\documentclass[12pt,preprint]{aastex}
\usepackage{epsf}
\usepackage{graphicx}

\newcommand{\etal}{{et al.~}}

\newcommand{\masyr}{ \ {\rm{mas \ yr^{-1}}}\>}
\newcommand{\kms}{ \ {\rm{km \ s^{-1}}}\>}
\newcommand{\PM}{{\rm PM}} 
\newcommand{\chour}{^{\rm h}\>}
\newcommand{\cmin}{^{\rm m}\>}

\begin{document}

\pagenumbering{arabic}
\title{The Proper Motion of the Large Magellanic Cloud using \textit{HST}}
\author{Nitya Kallivayalil\altaffilmark{1}}
\affil{Harvard-Smithsonian Center for Astrophysics, 
60 Garden Street, Cambridge, MA 02138}
\altaffiltext{1}{nkalliva@cfa.harvard.edu}
\author{Roeland P. van der Marel\altaffilmark{2}}
\affil{Space Telescope Science Institute, 3700 San Martin Drive, Baltimore, 
MD 21218}
\altaffiltext{2}{marel@stsci.edu}
\author{Charles Alcock\altaffilmark{3}}
\affil{Harvard-Smithsonian Center for Astrophysics, 60 Garden Street, Cambridge, 
MA 02138}
\altaffiltext{3}{calcock@cfa.harvard.edu}
\author{Tim Axelrod\altaffilmark{4}}
\affil{Steward Observatory, University of Arizona, Tucson, AZ 85721}
\altaffiltext{4}{taxelrod@as.arizona.edu}
\author{Kem Cook\altaffilmark{5}}
\affil{Lawrence Livermore National Laboratory, University of California, 
P.O. Box 808, L-413, Livermore, CA 94551}
\altaffiltext{5}{kcook@guen.ucllnl.org}
\author{A.~J.~ Drake\altaffilmark{6}}
\affil{Caltech, CACR, MC 158-79, 1200 E. California Blvd, CA 91125}
\altaffiltext{6}{ajd@cacr.caltech.edu}
\and
\author{M.\ Geha\altaffilmark{7,8}}
\affil{The Observatories of the Carnegie Institute of Washington,
    813 Santa Barbara Street, Pasadena, CA~91101}
\altaffiltext{7}{mgeha@ociw.edu}
\altaffiltext{8}{Hubble Fellow}

\begin{abstract}
We present a measurement of the systemic proper motion of the Large Magellanic Cloud (LMC) 
from astrometry with the High Resolution Camera (HRC) of the Advanced 
Camera for Surveys (ACS) on the Hubble Space Telescope ($HST$). We observed 
LMC fields centered on 
21 background QSOs that were discovered from their optical variability in 
the MACHO database. The QSOs are distributed homogeneously behind the 
central few degrees of the LMC. With  
2 epochs of HRC data and a $\sim2$ year baseline we determine the 
proper motion of the LMC to better than $5$\% accuracy:
  $\mu_W = -2.03 \pm 0.08 \ \rm{mas \ yr^{-1}}, \;
   \mu_N =  0.44 \pm 0.05 \ \rm{mas \ yr^{-1}}$. This is the most
 accurate proper motion measurement for any Milky Way satellite thus far. 
When combined with HI data from the Magellanic Stream this should 
provide new constraints on both the mass distribution of the Galactic Halo and  
models of the Stream.
\end{abstract}

\section{Introduction}
The Magellanic Clouds (MCs) are the closest, easily
observable galaxies from our vantage point in the Milky Way. They are
of fundamental importance for studies of stellar populations, the
interstellar medium, and the cosmological distance scale. Moreover,
due to their locations at 50 kpc (LMC) and 61 kpc (SMC) from the sun, and
$\sim 25$ kpc from the Galactic Plane, they provide one of our best
probes of the composition and properties of the Galactic dark halo.
 Precise measurements for the 
proper motions (PMs) of the
Magellanic Clouds combined with observational knowledge of the
distribution and velocities of HI gas in the Magellanic Stream  
can constrain models of the shape and radial density 
distribution of the
Galactic dark halo as well as theoretical models for the formation of
the Magellanic Stream.

Few developments in modern astronomy are as important
as the discovery of cosmic dark matter. Most of the matter in the
Universe is invisible to astronomers. In the Milky Way, the dark
component is about twenty times more massive than the visible disk of
stars and gas. The dark matter forms a vast, diffuse halo that
occupies more than a thousand times the volume of the stellar
disk. The composition of this dark halo is unknown, but it may
comprise a mixture of ancient degenerate dwarf stars and exotic,
hypothetical elementary particles (for a review see Alcock 2000).

The investigation of the Galactic halo is complicated by our immersion
deep inside it. Interior to the Solar circle the rotation curve may be
measured with confidence (see e.g. Fich \& Tremaine 1991). While the
estimates out to 20 kpc are not as robust, these observations 
indicate that the rotation curve is flat (Merrifield 1992; Yoshiaki \& Rubin 
2001). Much has
been learned about the halo in the direction out of the plane from the
study of the motions of gas clouds and stars. These imply that the
halo is more spherical than disk-like, but no precise measurement for
its axial ratio exists (van der Marel 2001). Analyses of the
kinematics of some globular clusters and small satellite galaxies
suggest that the halo extends out to $\sim 200$ kpc (e.g., Wilkinson
\& Evans 1999). For the large majority of these objects there is only
radial velocity data, and where available, proper motions are much
less precisely determined than the radial velocities. So while these
studies constrain the total halo mass, they say little about its shape
or distribution. Spherical symmetry and a particular density
distribution are generally {\it assumed}.

The Magellanic Stream is an HI emission feature that
spans more than $100^{\circ}$ across the sky (Br{\" u}ns \etal 2005). 
It consists of gas that
trails the Magellanic Clouds as they orbit the Milky Way and provides an 
 opportunity for detailed study of tidal disruption of galaxies as well 
       as the Milky Way dark halo. Many
detailed theoretical models have been constructed for the Magellanic
Stream (Gardiner \& Noguchi 1996; Lin, Jones \& Klemola 1995; 
Heller \& Rohlfs 1994). 
These models describe how the Magellanic Clouds orbit the
Milky Way, and how they lose material through tidal effects and other
physical mechanisms. The model parameters are adjusted to best
reproduce the observed density, morphology and velocities of the HI
gas seen along the Magellanic Stream. The most 
sophisticated calculations in the class of models that invoke tidal 
stripping are those by Gardiner \& Noguchi (1996). In their models the 
LMC and SMC form a
gravitationally bound system that orbits the Milky Way. The Magellanic
Stream represents material that was stripped from the SMC $\sim 1.5$
Gyr ago. This was the time of the previous perigalactic passage, which
coincided with a close encounter between the Clouds.

In Magellanic Stream models, the orbit of the clouds is one of the
prime variables that is adjusted to fit the data. As a result of
Newton's Law, the orbit is fully determined by: (a) the position of
the Clouds on the sky; (b) the present distance of the Clouds from the
sun; (c) the radial velocities of the Clouds; (d) the proper
motions of the Clouds; and (e) the gravitational potential of the
Galactic dark halo. The first three of these are known relatively
well; the latter two are not. Even small differences in the proper
motions can give vastly different orbits for the Clouds, especially when
considered over the long period of time over which the Magellanic
Stream developed. The usual approach has therefore been to estimate
the proper motions of the Clouds from the properties of the Magellanic
Stream, assuming a fixed Galactic halo gravitational potential
(usually a simple spherical isothermal halo). The results imply that
the Magellanic Clouds are just past their pericenter (itself at $\sim
45$ kpc from the Galactic Center), that the apocenter to pericenter
ratio is $\sim 2.5:1$, and the orbital period is $\sim 1.5$ Gyr. The
inferred tangential velocity in the Galactocentric rest-frame has
differed substantially from model to model, ranging from, e.g., 
$v_{\rm LMC,tan} =352$ km/s (Heller \& Rohlfs 1994) to $285$ km/s 
(Gardiner \& Noguchi 1996).

It was realized by Heller \& Rohlfs (1994) and Lin,
Jones \& Klemola (1995) that the arguments used to model the
Magellanic Stream can be turned around. If the proper motions of the
Magellanic Clouds are known, then the gravitational potential of the
Galactic dark halo can be determined from models of the Magellanic
Stream. They explored this using the available proper motion data, but
found that its accuracy was insufficient to obtain strong constraints
on the dark halo properties. Kroupa \etal (1994) found that a
proper motion better than $0.1\masyr$ is required
for strong results. 


So far the proper motion of 
the LMC has not been known accurately enough to strongly constrain 
properties of the Milky Way dark halo. Measurements, however, are
 steadily improving and determinations are available from
the following sources: Kroupa \etal (1994), using stars from the PPM
Catalogue; Jones \etal (1994), using photographic plates with a 14
year epoch span; Kroupa \& Bastian (1997), using Hipparcos data; Drake
et al.~(2002), using data from the MACHO project;  Anguita, Loyola
\& Pedreros (2000) and Pedreros et al.~(2002), using CCD frames with
an 11 year epoch span; and Momany \& Zaggia (2005), using the 
USNO CCD Astrograph all-sky Catalog (UCAC2). The measurements are all 
consistent with each other to within the error bars with the exception of 
two outliers: Anguita \etal (2000) and Momany \& Zaggia (2005)  
present values of $(\mu_{RA}, \mu_{DEC}) \sim (+0.84,+4.32)\masyr$. 
When these two results are ignored, the
weighted average of the remaining measurements yields proper motions
towards the West and North of $(\mu_W,\mu_N) = (-1.68 \pm 0.16, 0.34
\pm 0.16)\masyr$  (Van der Marel \etal 2002) i.e. to  $\sim 13$\%
accuracy.  This implies $v_{\rm LMC,tan} = 281
\pm 41$ km/s, which is consistent with most of the published Magellanic 
Stream models,
but is not accurate enough to discriminate between them. 
The proper motion of the SMC is much less well known than
that of the LMC, and only one reasonably accurate measurement exists.  
Kroupa \& Bastian (1997)  
obtain a value of $(\mu_\alpha cos(\delta),\mu_\delta) = 
(1.23, -1.21)\masyr$ 
for the SMC with an error of (0.84, 0.75)$\masyr$. 

A sound measurement of the proper motion of the LMC 
requires all of the following key
factors: (1) an instrument which can perform the astrometry with
adequate precision; (2) a reference frame consisting of point sources,
such as quasars, distributed widely behind the Clouds; (3) secure
determination of the membership of stars in the LMC; and (4) a
reliable kinematic model of the internal rotation of the Clouds. 
The previous estimates of the motions (referred to above) do not 
satisfy all of these requirements. However, a much improved measurement is 
now possible.  Some teams 
have identified QSOs behind our neighboring galaxies 
in order to provide good inertial reference frames against 
which the motions of these galaxies can be measured (Geha \etal 2003; 
Dobrzycki \etal 2002; 2003);
the $HST$ has been shown to be 
very stable for astrometry (Anderson \& King 2003) and in particular the  
Advanced Camera for Surveys (ACS) has higher resolution, and is better
 calibrated and more stable than 
even WFPC2 (Anderson \& King 2004; hereafter AK04).

We were awarded 
two epochs on the High Resolution Camera (HRC) on the ACS
 in Cycle 11 and Cycle 13 for a study of the proper motions of the MCs. 
We report our proper motion result for the LMC in this paper. SMC results 
as well as implications 
for the MC system as a whole will be presented in  subsequent papers. The paper is 
outlined as follows: \S2 deals with the sample and the observations; \S3 
describes our analysis; \S4 presents the center of mass proper motion 
of the LMC; \S5 is a discussion and \S6 is a summary.

\section{Sample Selection and Description of Observations}
\subsection{Sample}
A total of 54 QSOs were identified behind the Magellanic Clouds from their optical
variability in the MACHO database (Geha et al. 2003). Out of 
this sample, 44 QSOs were behind the LMC.
In our first epoch program (GO-9462; Cycle 11), we proposed to image
fields around these QSOs  with ACS/HRC in
snapshot mode.  In a snapshot proposal the targets are 
randomly selected from the proposed list, but with manually assigned 
priorities. We successfully observed 34 of the 44 LMC QSOs in the first epoch. 
In the second epoch program GO-10130 (Cycle 13) we  
re-imaged the targets that were observed in epoch 1 and for
maximally efficient use of $HST$ resources we again chose to observe in 
 snapshot mode. We achieved a completion rate of $62$\% for epoch 2  
and $48$\% overall, so that our final sample consists of 21 targets. 
Table~1 describes this sample. It lists the ID number for 
each QSO (each QSO field will be referred to by its ID number 
subsequently), as well as its MACHO ID (for referencing purposes 
with Geha \etal 2003), RA, DEC (J2000), $V$ magnitude (an average of 
our measurements from both epochs) and redshift.
Table 1 also lists the execution date, visit number, exposure time and 
orientation for each epoch's observations, as well as the time baseline 
that was ultimately achieved for each target. These are discussed in 
\S2.2.

\begin{figure}
\plotone{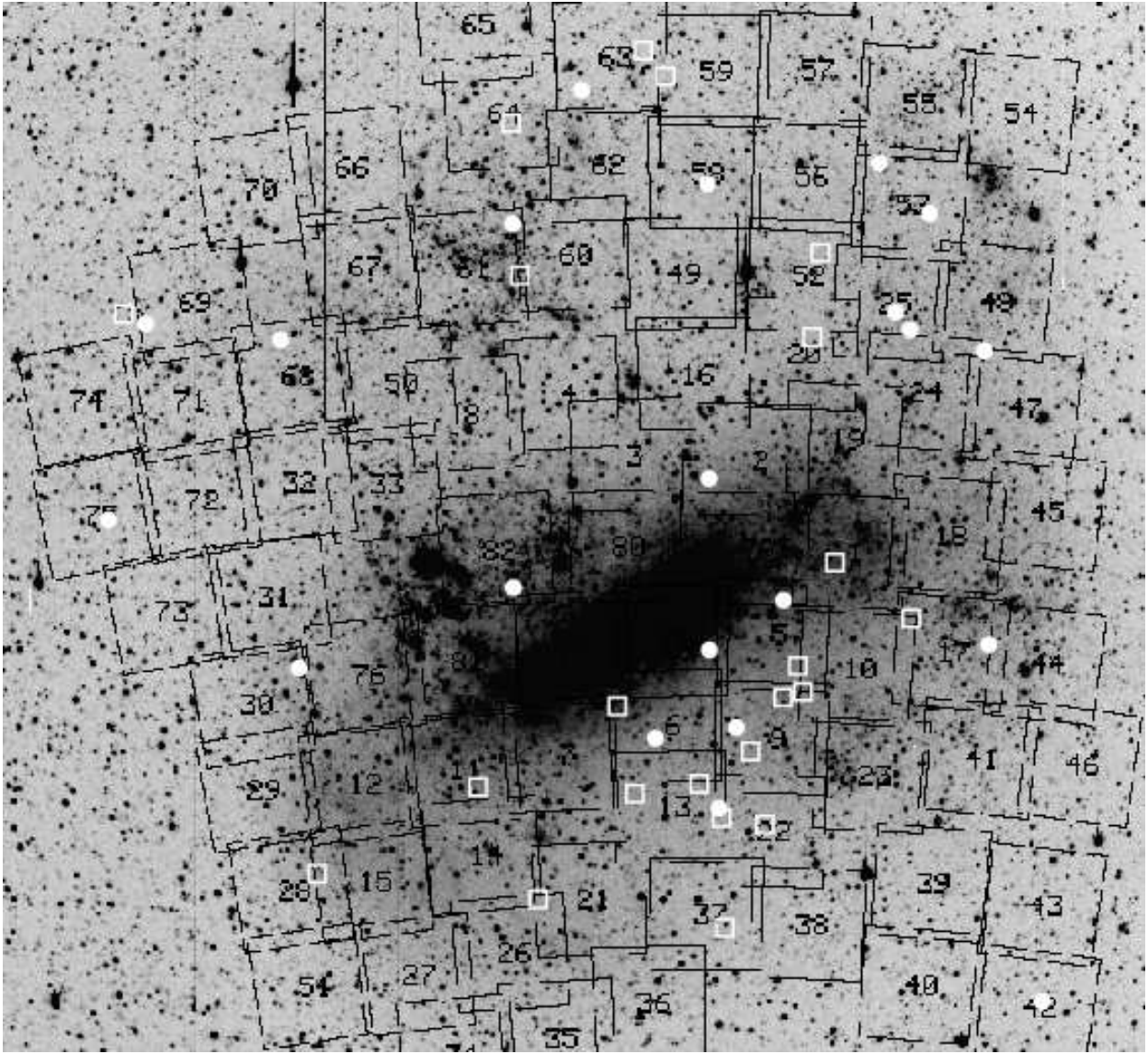}
\caption{R-band image of the LMC ($8^{\circ} \times 8^{\circ}$). 
The MACHO photometric coverage
is indicated. White circles indicate reference QSOs for which we
have obtained two epochs of ACS/HRC imaging and which we subsequently 
use in the PM analysis; squares indicate QSOs which we did propose for 
but for which we did not get two epochs of imaging in our snapshot program.}
\label{figure1}
\end{figure}

Figure~\ref{figure1} is a plot of the QSOs behind the LMC. QSOs for 
which we obtained two epochs of data are marked with white circles. Their 
distribution behind the LMC is reasonably uniform. QSOs which did not 
make it through one or both of our snapshot programs are marked with white 
squares.

\subsection{Scheduling and Execution}

\begin{figure}
\plotone{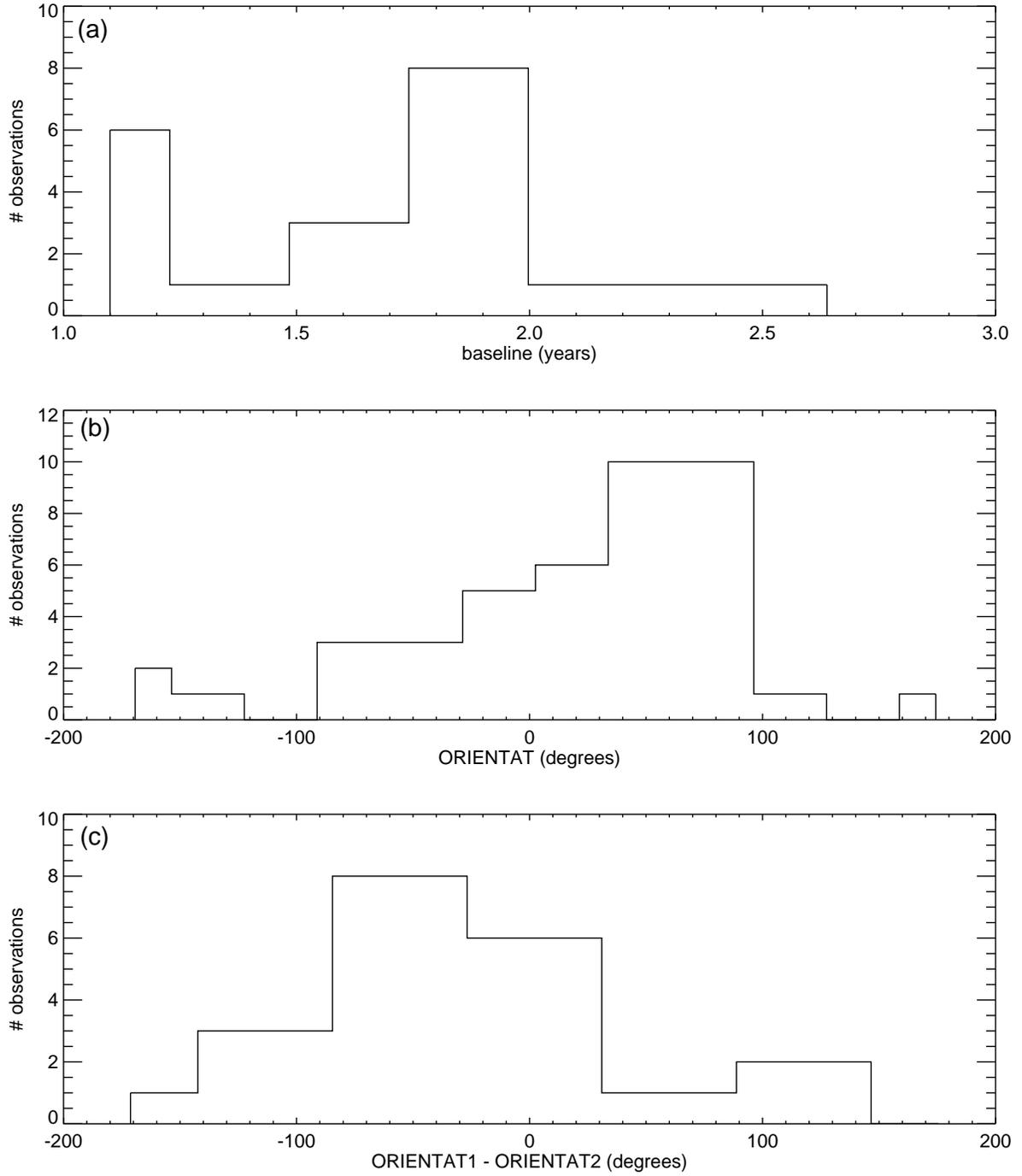}
\caption{Histograms of 
(a) 
number of QSO fields with  a given 
time baseline; (b) number of QSO fields taken with 
 a given telescope orientation on the plane of the sky (ORIENTAT); and 
(c) number of fields at a given
 relative telescope rotation between the two epochs.}
\label{figure2}
\end{figure}

In a snapshot program the targets are executed in the 
cycle so as to fill gaps in the $HST$ schedule. The 
majority of our observations were 
scheduled in the first part of the cycle. It is interesting to study some 
of the specifics of the scheduling times and the quantities that 
they affect.  Figure~\ref{figure2}(a) is a histogram of the 
time baseline for each QSO field (in years). The median baseline that 
we achieved is $\sim1.9$ years  but there is a spread from $\sim1.1$ 
to 2.7 years. For observations with $HST$, the date of observation 
determines the orientation of the detector with respect to the 
sky (see the $HST$ Primer, Karakla 2005). Again, because of the snapshot 
nature of our project, the  orientations were not fixed \textit{a priori} 
and hence were  not identical from one epoch to the next.
Figure~\ref{figure2}(b) is a histogram of the orientations on the sky 
(ORIENTAT) of each of our QSO fields 
for both epochs (ORIENTAT is defined as the angle on the sky, East of North, 
of the detector $y$-axis). 
Figure~\ref{figure2}(c) is a histogram of the  relative telescope rotation 
between the epochs for each QSO field, plotted between -180\degr and 
180\degr~. It shows that the observations were not taken at completely 
random orientations, for there is a broad peak near $\sim0$\degr. Nonetheless, 
there is considerable randomness in the orientation of the telescope 
with respect to the sky, both in an absolute and in a relative sense. This is 
good because it implies that any possible systematic proper motion errors 
tied to the CCD frame will average to zero roughly as  $1/\sqrt{N}$. 
The possibility of actual systematic effects are investigated in detail in \S 3.3. 

 Our observing strategy involved the use of the HRC because it
provides the highest resolution available  (the average pixel scale is
28.27 mas/pixel - see AK04)  and is well calibrated for astrometry. It
is also well sampled even in  the bluest filters and thus a star's
integrated flux does not depend  strongly  on where it lands in a
pixel. It is optimized towards the visual and  red part of the
spectrum and so for our main astrometry goals we chose the  F606W
filter which is a broad $V$ filter with high throughput. This filter is also
not  too red and thus avoids possible issues with the HRC red halo at
wavelengths $\gg7000$\AA \ (see the ACS Instrument Handbook, Pavlovksy \etal 
2005). In addition to F606W, we
used F814W which is a broad $I$ filter for imaging in the first epoch only, not
for the main astrometry,  but to be able to place sources on a
color-magnitude diagram (CMD). We thought that this might help assist in the
rejection of foreground stars.

For the F606W filter, in epoch 1, exposure times were chosen 
so as to achieve a S/N of at least 100 for the QSOs based on the 
known MACHO magnitudes. In epoch 2 the exposure times were chosen to 
be somewhat larger than in epoch 1 to take into account the fact that 
the MACHO sources were considerably blended and that the actual $V$ 
magnitudes measured from first  epoch $HST$ data tended to be dimmer than 
the MACHO magnitudes. 
Some of the isolated QSOs actually turned out to be brighter 
because of their intrinsic variability. Note that on the timescale of our 
proper motion measurements, the expected photometric variability is only  
approximately a few tenths of a magnitude in the $V$-band and so we quote 
an average from both epochs in Table~1. 
For the F814W filter, the exposure time was set to 1.7 minutes total which is 
sufficient for photometry.

\begin{figure}
\plotone{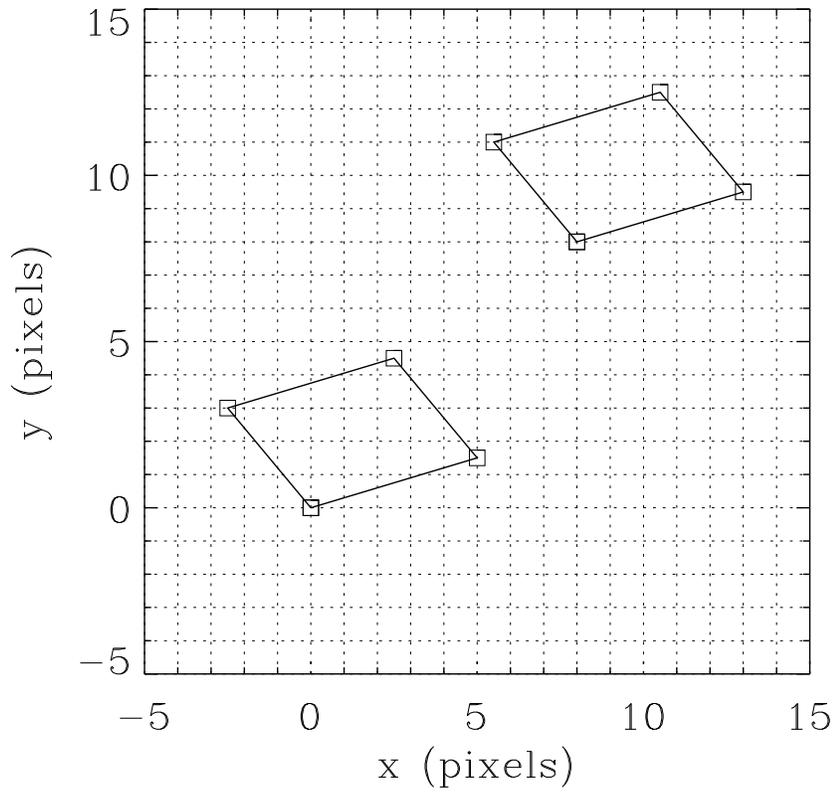}
\caption{The dither pattern used for 
the F606W images.}
\label{dithers}
\end{figure}

We chose the same dither pattern in the F606W filter for both epochs.
This amounted to 8 exposures per epoch for each QSO (16 exposures
total for both epochs) taken using two four-point half-pixel shifts
which were shifted relatively by an integer 8-pixel amount. 
 Figure~\ref{dithers} is a schematic of this dither pattern. Since we
did not plan to do astrometry with the F814W images we chose a simple
CR-SPLIT in order to be able to reject cosmic rays and other
transients. In summary, each QSO field has 18 images 
associated with it - 16 in $V$-band and 2 in $I$-band - and each of the 
$V$-band images were dithered to minimize any pixel-location based 
systematic errors.

For our basic reduction purposes we used the bias-subtracted,
dark-subtracted, flat-fielded images (\_flt.fits) provided by the
STScI/ACS data reduction pipeline.  AK04 found that these images are
well-behaved for astrometric purposes. They are not, however,
corrected for geometric distortion.  We did not use the geometrically
corrected products created by the MultiDrizzle software (\_drz.fits) since they
involve a re-sampling that might degrade the astrometry.  Instead of
geometrically correcting the images, we geometrically corrected the star
positions on the images, after doing PSF-fitting, as we will
discuss in \S 3 below.

\section{Analysis}

\subsection{Methodology for PSF-fitting and Distortion Solution}
 The HRC has a large amount of geometric distortion due to
its off-axis location and it is necessary to understand this fully
because a high-precision astrometric program such as ours requires a
distortion solution that is at least as accurate as the centroiding.
Although the distortion is large, it is very stable, and can therefore
be calibrated accurately.  This calibration has been done by AK04, 
and our analysis relies heavily on their work. They
and Krist (2003) have shown that for ACS/HRC the Point Spread Function
(PSF) doesn't vary significantly over the field (in contrast to
WFPC2).  Although the $HST$ PSF does vary with time due to
``breathing'' (thermal effects associated with day/night transitions)
and long-term changes in focus, these changes affect mostly the
relative amounts of light in the core and the wings of the PSF. The
actual shape of the core is not altered significantly. AK04 developed
multi-filter PSFs for the HRC using a large dataset of 40 observations
each with 20 different pointings of the globular cluster 47 Tuc ($HST$
Calibration Programs GO-9028, GO-9443 \& GO-9019).

AK04 also developed a detailed solution to the geometric distortion
 which they modeled as the combination of a large-scale distortion
 characterized by a fourth-order polynomial and fine-scale variations
 characterized by a look-up table that can be linearly interpolated to
 give the adjustment required at any point in the image.  They showed 
 that the second and higher order terms are stable over short and long
 timescales. However, the linear terms vary with time (even within a
 single $HST$ orbit) due to breathing, differential velocity
 aberration, and other effects.  Therefore any two exposures must be
 brought to a common frame with the use of a 6 parameter linear
 transformation (the 6 parameters are scale, skew, translation and
 rotation).  When this is done, the geometric distortion correction is
 accurate and stable over long timescales to $\sim$0.005 pixel.  For
 comparison, given our median time baseline (see Figure~\ref{figure2}), 
the proper
 motion of the LMC, estimated on the basis of previous work (Van der
 Marel \etal 2002), corresponds to a position difference of 0.1 pixels
 between epoch 1 and 2. Thus we have a very good handle on
 systematics in this regard.

To sum up, AK04 have provided software that finds the point sources on
 an image, fits the PSF to determine the raw stellar position and
 flux, applies a distortion correction to calculate the geometrically
 corrected stellar position (i.e., on a frame that is locally
 Cartesian on the sky) and finally, applies a photometric correction
 to the sources for geometric distortion and finite aperture effects.

\subsection{Data Analysis}
We ran the AK04 code on all 18 images for each field.  At this stage, 
some cuts were made based on
crowding (by specifying how far away the next brightest source can be) and
threshold (by specifying the minimum peak brightness in excess 
of the sky background). These cuts were not stringent so as to 
identify the largest possible number of real sources. Using the first $V$-band 
image in the first epoch
as a reference, we cross-identified the detected sources in all other
images with it.  We then created a master-list of sources for each QSO
field by accepting only those sources detected in all 18 images and
for which the magnitudes in all $V$ images were in mutual agreement to
within $6\sigma$. This rejected all cosmic rays, other transients and
bad pixels that were detected as sources (note that individual
exposures were short so that cosmic ray rejection would be reliable).  
This also rejected all bona fide stars that had fallen off
the field in just a few of the images because of the dither pattern
or the CCD rotation between epochs 1 \& 2, 
and stars in which one or more of the measurements were 
sub-optimal. So in this respect this cut was conservative. 
However, it ensured
that we used exactly the same subset of stars for every image in every
epoch, which in turn minimized any potential for systematic
differences and errors.  Table 2 lists for each field the number of
real sources (detected in at least half of the images) and the number
of sources in the master-list (detected in every image in every epoch
with magnitudes consistent to within $6\sigma$). As can be seen from
the table, this generally cuts down the number of sources by a factor of two.

 We obtained an average $V$ and $I$ magnitude and photometric errors
for each source in the master-list by combining the measurements of
the individual exposures.  For each field we identified the QSO in the
master-list from registering
with the MACHO discovery images (see Geha \etal 2003). We then removed the 
QSO from the master-list to obtain a list that consisted primarily of
LMC objects.
We were not interested in measuring (nor were we able to measure) the
 relative motion of LMC stars within a given field with respect to
 each other.  We could only measure the average motion of all the stars
 in a given field with respect to the QSO. To do this we determined 
and applied
 geometric transformations to align the star positions in every image
 with those of the reference image.  Then subsequently, we used 
the same transformations to correct the position of the QSO. The 
motion of the QSO, over the time baseline, was then obtained simply 
as the difference between its
 average position in each epoch. Since in reality the QSO is so
 distant that it is essentially fixed on the sky, the resulting
 measurement is just the reflex motion due to the average motion of
 LMC stars.  

We used a six parameter
 linear fit to transform the star positions and align them as best as
 possible (in a $\chi^{2}$ sense) with the positions of the same stars
 in the reference image.  We ran the linear transformations
 iteratively and accepted for use in the calculation of the
 transformation only those stars for which the proper motion (PM) and
 the error in the proper motion ($\delta$PM), measured relative to the
 average of the other stars, were both $<0.1$ pixels. This rejected
 foreground stars (in general those are expected to move by
 $\sim8\masyr$ towards the North; see Momany \& Zaggia 2005) and 
stars with large centroiding errors. The latter are 
generally faint stars that were unlikely to
 improve the accuracy of the result. The terms in the
 transformations were modified iteratively until the number of stars
 used in the fit was unchanging.

\begin{figure}
\plotone{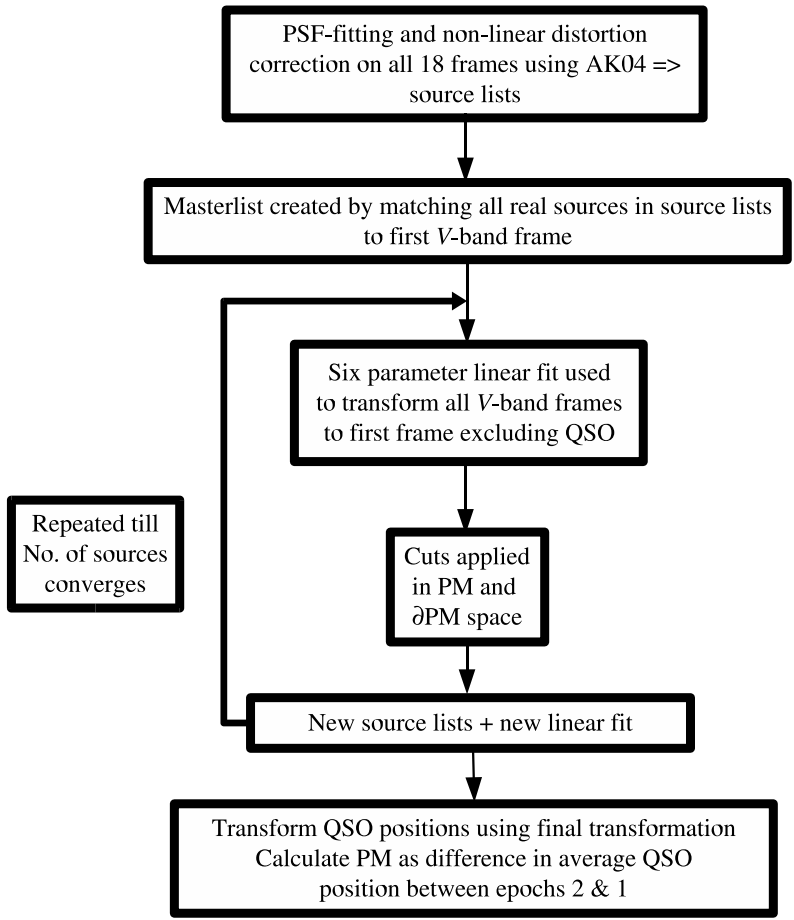}
\caption{A flowchart outlining the steps involved in getting a
PM (in pixels) for each QSO field.} 
\label{flowchart}
\end{figure}

\begin{figure}
\plotone{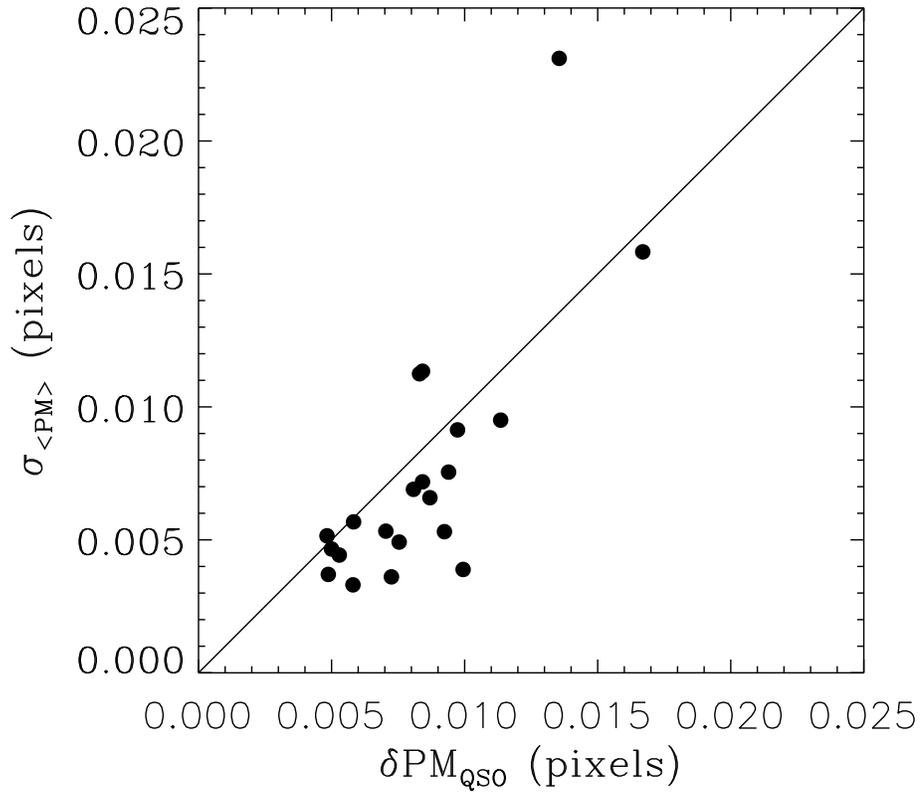}
\caption{The distribution of errors in our linear transformations for each 
field. The $x$-axis shows the error in the PM of the QSO and the 
$y$-axis shows the error in the average PM of the star-field.}
\label{figure3}
\end{figure}

After determination of the linear transformations for all images, we
calculated the PM and $\delta$PM for all sources in the master-list,
as well as for the QSO.  Figure~\ref{flowchart} is a 
schematic illustrating the basic steps involved in getting a QSO PM in pixels.
 One component of the final astrometric error
is how well we are able to align the stars between epochs.  We
therefore determined the PM of each star individually as the
difference between its average positions in epoch 2 and 1 (just as we
do for the QSO).  We then ran some statistics to calculate the error
in the average PM of all the stars, which we called
$\sigma_{\rm{\langle PM \rangle}}$ (the average itself
is zero by definition).  This error quantifies how accurately we were
able to align the star-fields between the two epochs and gives us an
idea of our systematic errors. Table 2 lists the final number of 
sources ($N_{\rm{used}}$) 
used in the linear transformations for each field.  Once the
images are aligned the relative motion of the QSO and the
star-field is simply the difference in the average position of the QSO
from epoch 1 to epoch 2. We called the random error in this difference 
$\delta \rm{PM_{QSO}}$. The final error in the relative motion
between the QSO and the star-field is then $\sqrt{\delta \rm{PM_{QSO}}^{2}
+ \sigma_{\rm{\langle PM \rangle}}^{2}}$.  Figure~\ref{figure3} shows the
distribution and contributions of these two components of the final
error.  $\delta \rm{PM_{QSO}}$ is plotted on the $x$-axis and
$\sigma_{\rm{\langle PM \rangle}}$ is on the $y$-axis. The straight line
marks $\delta \rm{PM_{QSO}}=\sigma_{\rm{\langle PM \rangle}}$.  The
figure shows that our final errors are dominated by the centroiding
errors of the QSO and that the error introduced by aligning the
star-field is smaller.

We transformed the results to a PM and $\delta$PM of LMC stars in the
directions North and West in $\masyr$ using the orientation of the
reference image with respect to the sky (ORIENTAT), the time baseline,
and the fact that the PM of the LMC is the opposite of what we 
measure for the QSO (technically what we measure is the reflex motion
of the QSO). Absolute orientations of $HST$ observations are fairly accurate
($\la 0.003$ \degr) and any uncertainty in ORIENTAT has no effect on our 
final PM errors.  Table 2 tabulates the PM for each field and the
corresponding error (columns 5 - 8).

\subsection{Inspection and Consistency Checks}
\begin{figure}
\plotone{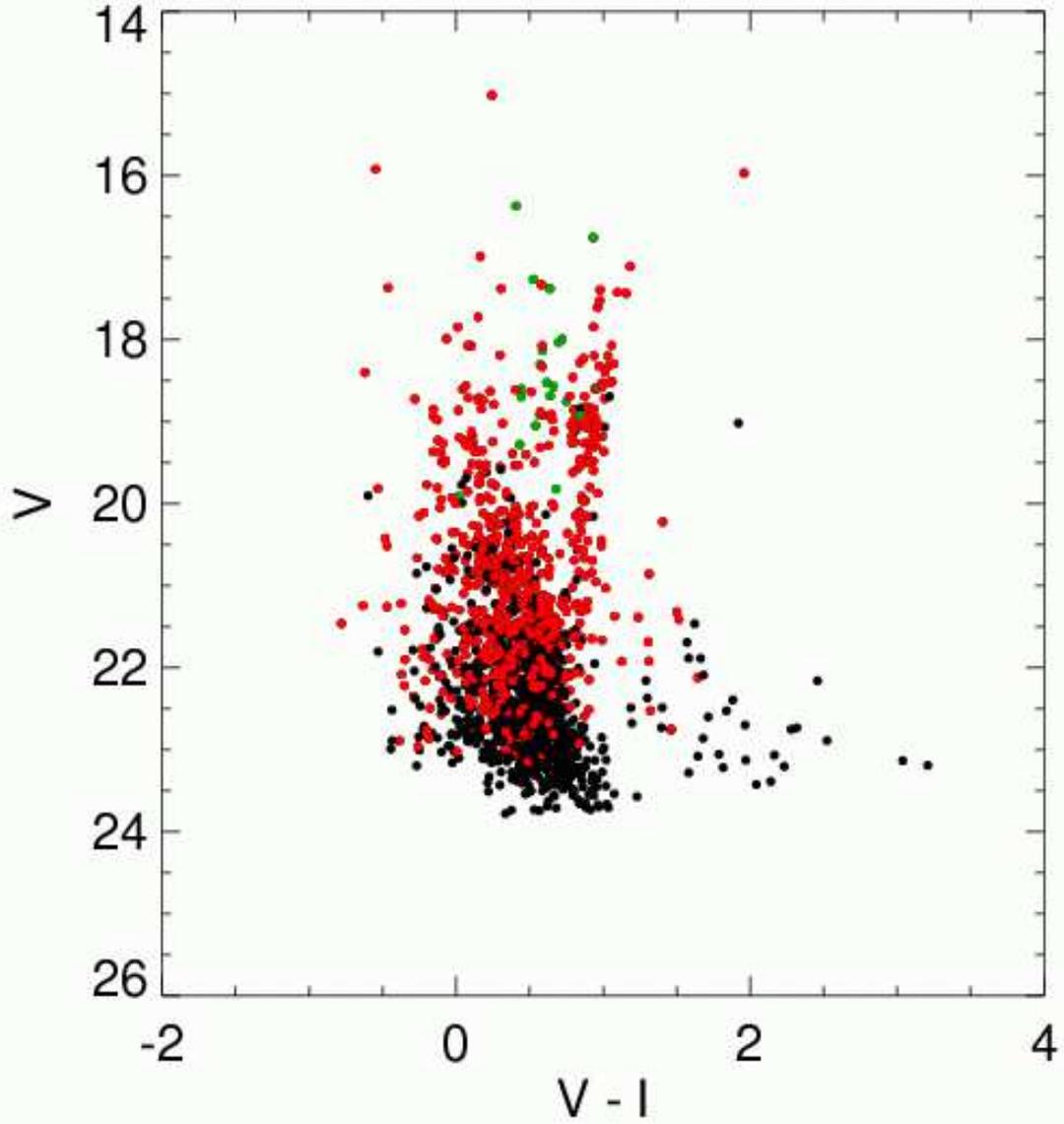}
\caption{ ($V-I$, $V$) color-magnitude diagram for the LMC region of the sky.
QSOs are marked in green, stars in the master-list with PM \& $\delta$PM
 $<0.1$ pixels are 
marked in red and the rest are shown in black.}
\label{figure4}
\end{figure}

Figure~\ref{figure4} shows the ($V-I$, $V$) CMD for the LMC.  
The QSOs are marked
in green and the  stars with PM \& $\delta$PM $<0.1$ pixels are marked
in red.  The CMD looks consistent  with other CMDs of the LMC (see for
e.g. Alcock \etal 2000b). The  main sequence and the giant branch are
clearly visible, and the red clump  (which is often used as a distance
indicator) is also  where we would expect it to be,  at $\sim V=19.1,
\ V-I \sim 1$ (see Smecker-Hane \etal
2002). We believed that we might make some additional cuts for a 
foreground population of stars based on this CMD, but such a population 
is not clearly delineated, and any cuts we could make (like the very red, 
dim tail of objects) seem to be obviated by the PM cuts that we employ.

To test the robustness of our results we  tried a number of different
variations and cuts in our analysis. Apart from the cuts in PM space
to eliminate foreground  stars and stars with large centroiding
errors, these
variations included:   a) constraints on the distance of a star from
the center of the field (i.e. eliminating  stars in the outer edges
from our analysis in case the geometric  distortion is more
complicated than understood in those regions); b) inclusion of higher
order terms in the transformation to   account for any possible
time-variations in the  distortion terms; c) making both steeper  and
looser cuts in the PM and $\delta$PM space of the  stars (e.g. PM and
$\delta$PM $<0.05$ and $0.2$ pixel); d) using only those stars whose
PM/$\delta$PM $\le2.5$ to further ensure that the transformations 
are not unduly influenced by stars whose PMs are not statistically 
consistent with zero. None of 
these cuts were found to  significantly alter the  results. In
particular they did not affect the fields that could be seen as
`outliers' at that point in the analysis.

\begin{figure}
\plotone{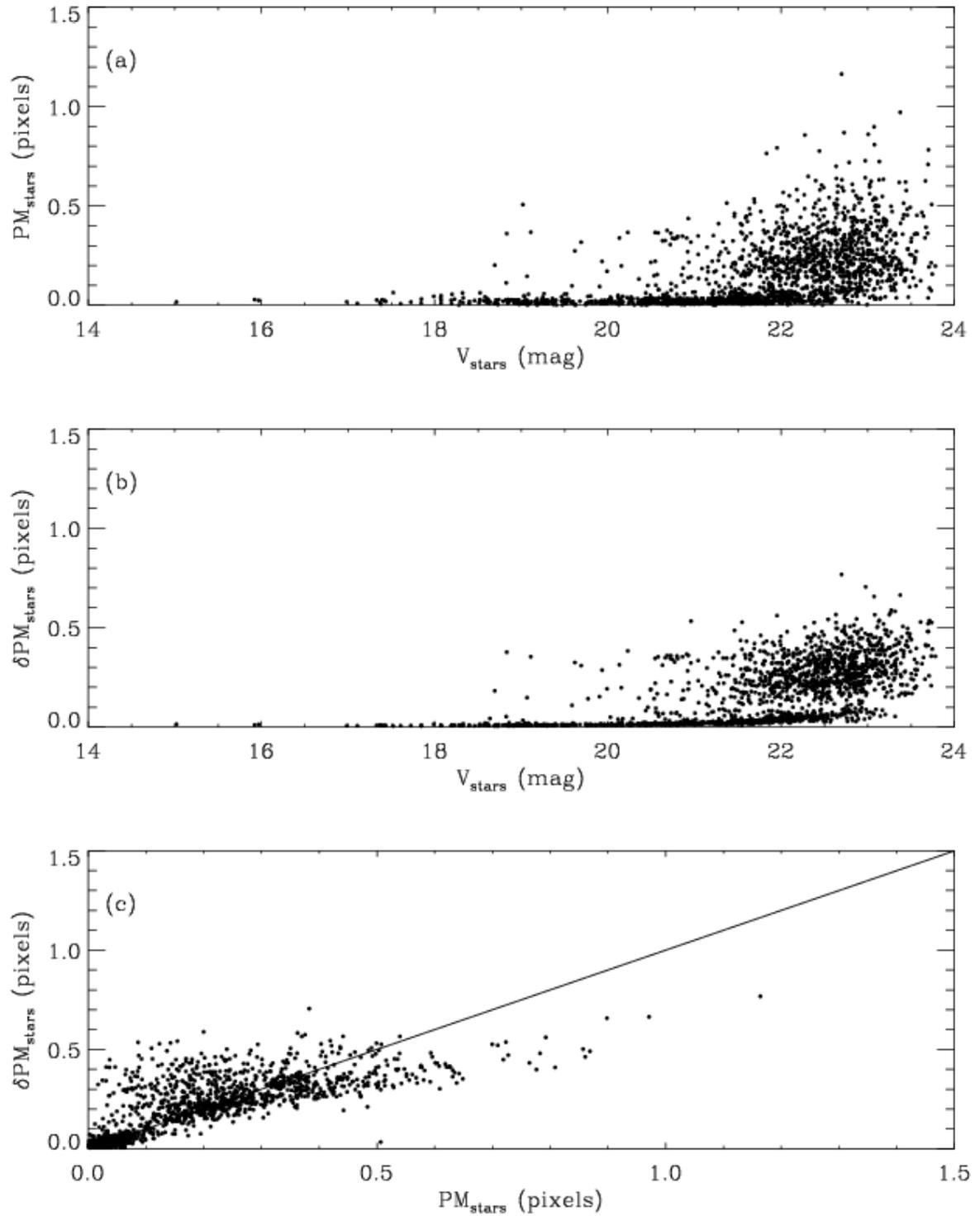}
\caption{(a) PMs for all stars in all master-lists 
as a function of their $V$ magnitude; (b) $\delta$PM for all stars in all 
master-lists as a function of $V$ magnitude; (c) $\delta$PM vs. PM for all 
stars in all master-lists.}
\label{systematics}
\end{figure}

Figure~\ref{systematics}(a) \& (b) show PM and $\delta$PM in pixels
as a function of $V$ magnitude for all stars in all master-lists of
all fields respectively. These plots 
look reasonable and have the expected shapes with errors $\propto
\rm{(S/N)}^{-1}$.   Figure~\ref{systematics}(c) shows $\delta$PM
vs. PM for all stars in all master-lists. It 
illustrates that the PM residuals are not much larger than the
expected random errors.

\begin{figure}
\plotone{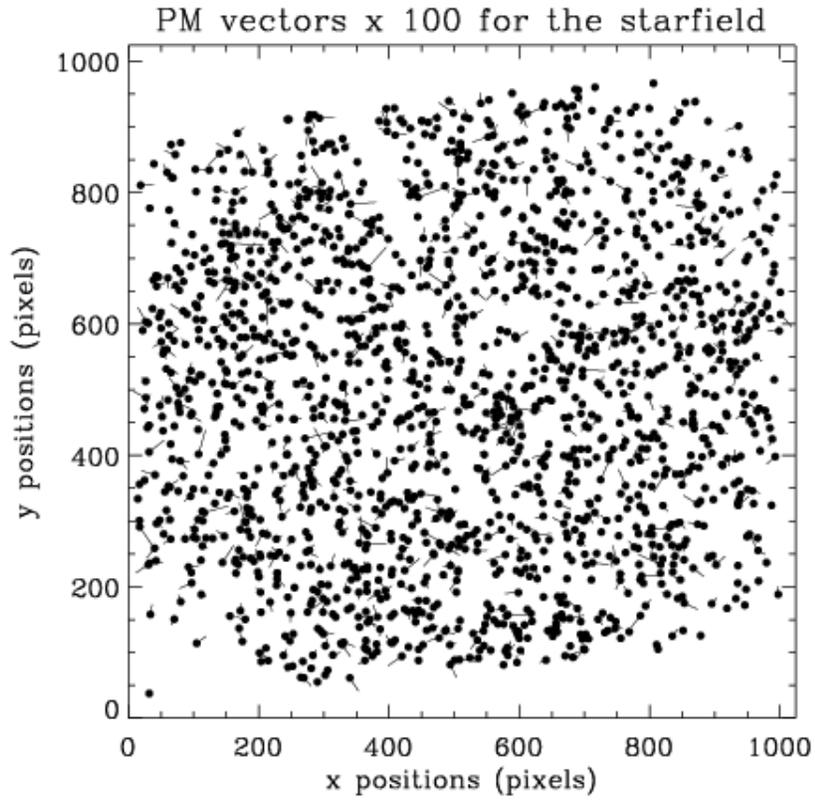}
\caption{PM vectors of all stars in all master-lists 
magnified by a factor of 100.} 
\label{pmvectors}
\end{figure}

\begin{figure}
\plotone{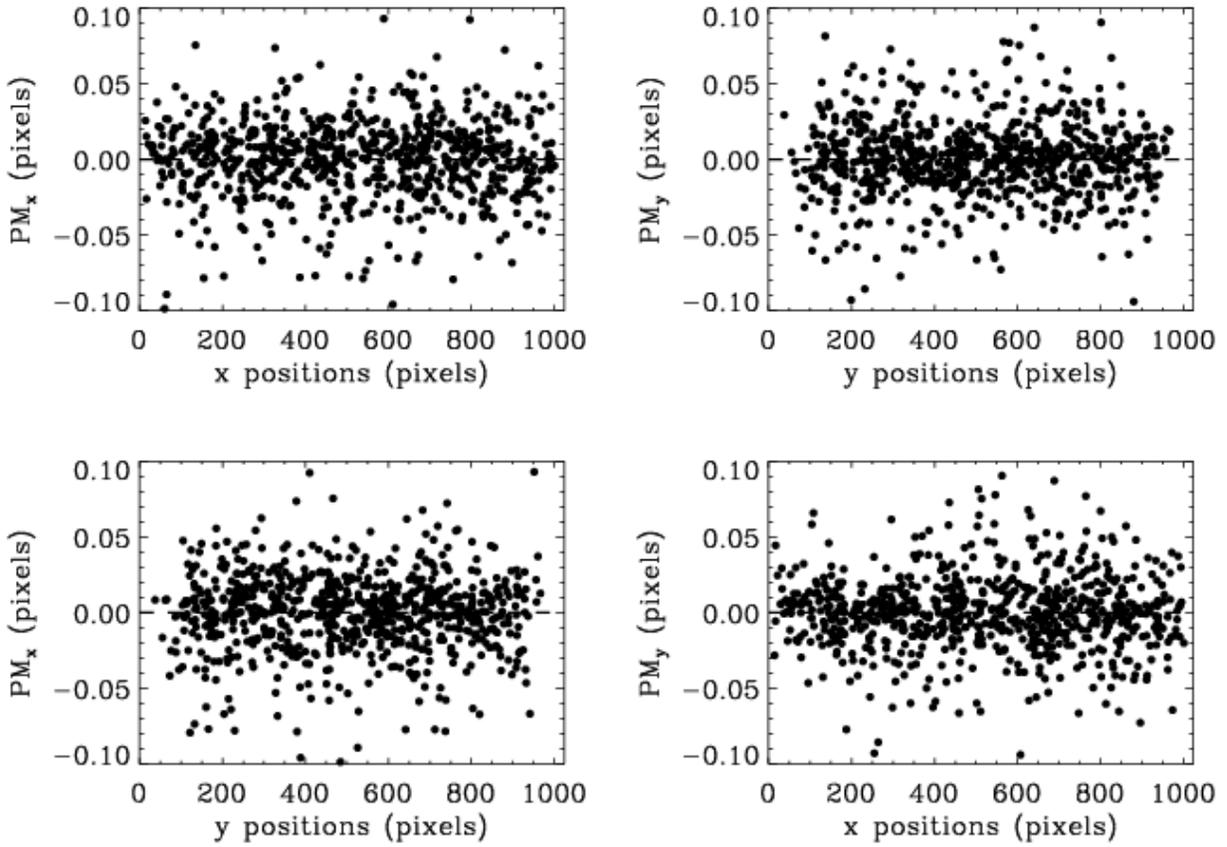}
\caption{PMs of 
the stars in the masterlist that have PM and $\delta \rm{PM} <0.1$ pixels 
versus chip location separately for $x$ and $y$ to 
see if there are any systematic trends with position.
These are the stars that are used in our final linear transformations.}
\label{pmvspos}
\end{figure}

\begin{figure}
\plotone{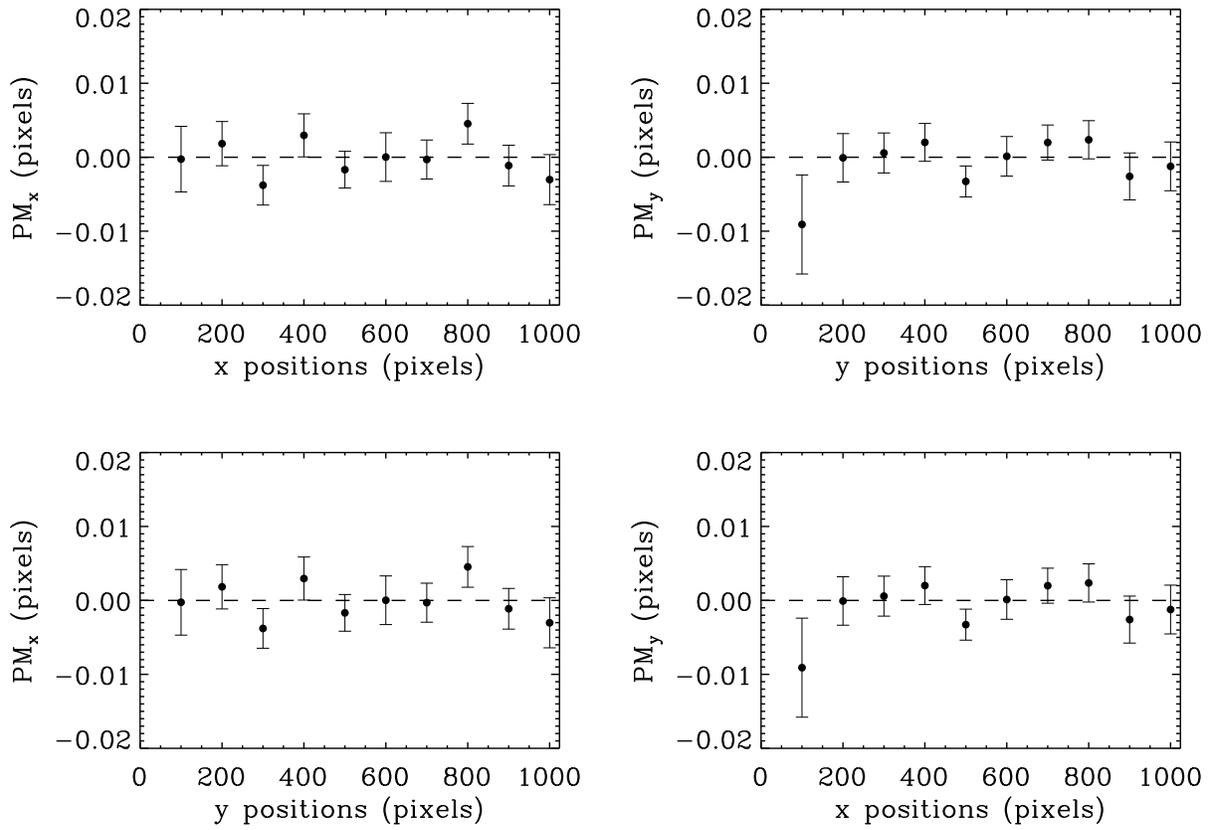}
\caption{Average PMs of 
the stars in the masterlist that have PM and $\delta \rm{PM} <0.1$ pixels 
versus chip location. The PMs of the stars have been binned for every 100 
pixels and then averaged.}
\label{pmbins}
\end{figure}

Figure~\ref{pmvectors} shows PM vectors for all the stars in all 
master-lists of all
fields in $x$ and $y$-space in pixels, magnified by a factor of
hundred. This enables us to inspect how and if the shifts vary with
position on the chip and if there are any other systematic trends in
the transformations. All vectors appear to be random both in magnitude
and orientation across the field. Figure~\ref{pmvspos} seeks to 
address the same concerns, but plots the PMs for the stars with 
PM and $\delta{\rm{PM}} < 0.1$ pixels versus 
$x$ and $y$ position on the chip separately. The scatter 
in $x$ and $y$ look comparable and there is no obvious 
trend with position on the chip. Figure~\ref{pmbins} is again 
a plot of PM vs. $x$ and $y$ for the stars, but now the 
PM values have been binned for every 100 pixels, and the average PM value for 
each bin is plotted. This allows us to really get down into the noise and 
investigate systematic errors at the level of a hundredth of a pixel. 
We conclude from this plot that there is no evidence for systematic
 errors in the geometric distortion correction larger than $\sim0.005$ pixels,
 which is consistent with the findings of AK04.
Thus, all the results
look sensible, robust, and well behaved to within the expected accuracy.

\section{Proper Motion results}

\subsection{Field Dependence of Proper Motions}

One way to estimate the proper motion of the LMC center of mass
 is just to average the results from the 21 fields listed
in Table 2. However, this is not the most accurate approach. The
fields are separated by several degrees and this causes real
variations in the PMs of LMC stars in different fields. In general,
the motion observed for an individual QSO field can be written as
\begin{equation}
  \PM({\rm field}) = \PM({\rm CM}) + \PM_{\rm res}({\rm field}) .
\end{equation}
Here $\PM({\rm CM})$ is the proper motion of the LMC center of mass and
$\PM_{\rm res}({\rm field})$ is the field-dependent residual. The
latter contains contributions from different effects:
\begin{description}
\item[(a)] variations as a function of position in the components of 
the three-dimensional velocity vector of the center of mass that are
seen in the plane of the sky (``viewing perspective'');
\item[(b)] the internal rotation of the LMC;
\item[(c)] time variations $di/dt$ and $d\Theta/dt$ in the LMC viewing angles.
\end{description}

The formulae for these contributions can be found in van der Marel
\etal (2002, hereafter vdM02). With the help of a model for these
contributions one can use the observed proper motion $\PM({\rm
field})$ for any given field to obtain an estimate $\PM_{\rm est}({\rm
CM})$ for the proper motion of the LMC center of mass using
\begin{equation}         
\label{correctionformula}
  \PM_{\rm est}({\rm CM}) \equiv \PM({\rm field}) - 
     \PM_{\rm res}({\rm field}) .
\end{equation}
The estimates $\PM_{\rm est}({\rm CM})$ for all fields are listed in
Table~2 (columns 9 \& 10). The field-dependent correction terms 
$\PM_{\rm res}({\rm field})$ are small compared to the overall LMC 
proper motion, but are
significant given the accuracy that we are trying to achieve.
The average two-dimensional size of the correction terms is $0.20
\masyr$ and the maximum over all fields is $ 0.53\masyr$. The
corrections are smaller for fields that are closer to the center of
the LMC. The largest contribution to the corrections comes from the
viewing perspective effect, which contributes on average $0.15 \masyr$,
with a maximum over all fields of $0.31 \masyr$. By contrast, the
internal rotation contributes on average $0.06 \masyr$, with a maximum
over all fields of $0.20 \masyr$.

We used an iterative procedure to evaluate 
equation~(\ref{correctionformula}). This is necessary because the
residuals $\PM_{\rm res}({\rm field})$ themselves depend either
directly or indirectly on the proper motion of the LMC center of mass.
At each iteration step we averaged the estimates $\PM_{\rm est}({\rm
CM})$ for the different fields to obtain an optimal estimate for 
$\PM({\rm CM})$. This value was then used in the next iteration step
to evaluate the residuals $\PM_{\rm res}({\rm field})$. For the
internal rotation model of the LMC we fitted the line-of-sight
velocity data for 1041 carbon stars (at each iteration step) as in
vdM02. This fit depends on the proper motion of the LMC center of
mass, because the latter introduces a spurious solid-body component in
the observed line-of-sight velocities that must be subtracted while
modeling the internal rotation. We did not include any time-dependence
of the viewing angles (indicative of precession and nutation of the
LMC disk plane) in the model. As we show in \S4.4 below, the
available observational constraints are consistent with $di/dt =
d\Theta/dt = 0$. Where necessary, in the model for $\PM_{\rm
res}({\rm field})$ we used: distance modulus $m-M = 18.50$, systemic
line-of-sight velocity $v_{\rm sys} = 262.2 \kms$, center position
$\alpha_{\rm CM} = 5\chour 27.6\cmin$ and $\delta_{\rm CM} =
-69.87^{\circ}$, inclination $i = 34.7^{\circ}$ and line-of-nodes
position angle $\Theta = 129.9^{\circ}$ as in vdM02.

The field-dependent corrections $\PM_{\rm res}({\rm field})$ are
more-or-less symmetrical around the LMC center. 
So the corrections approximately average to zero for a
distribution of quasar fields that is symmetric around the LMC
center. This is not exactly the case for the fields that we observed, but
nonetheless, the fields are reasonably homogeneously distributed around the
center (see Figure~\ref{figure1}). Therefore, the average of the 
corrections over all fields is 
smaller than the individual corrections 
themselves. Moreover, this average is determined primarily by the
geometry of the field distribution. The specific assumptions made
about the model of the LMC and its rotation have almost a negligible
effect. We have run tests in which we varied the values of the model 
parameters over ranges that are realistic estimates of the
observational uncertainties: $5 \kms$ in $v_{\rm sys}$,
$10^{\circ}$ in inclination, $120 - 155^{\circ}$ for the 
range in line-of-nodes position
angle, $0.5^{\circ}$ in LMC center position, and $0.1$ mag in distance
modulus. None of these changes affected our final estimate of the
proper motion of the LMC center of mass by more than $0.02 \masyr$.

\subsection{Check for Systematic Errors}
  
\begin{figure}
\centerline{
\epsfxsize=0.9\hsize
\epsfbox{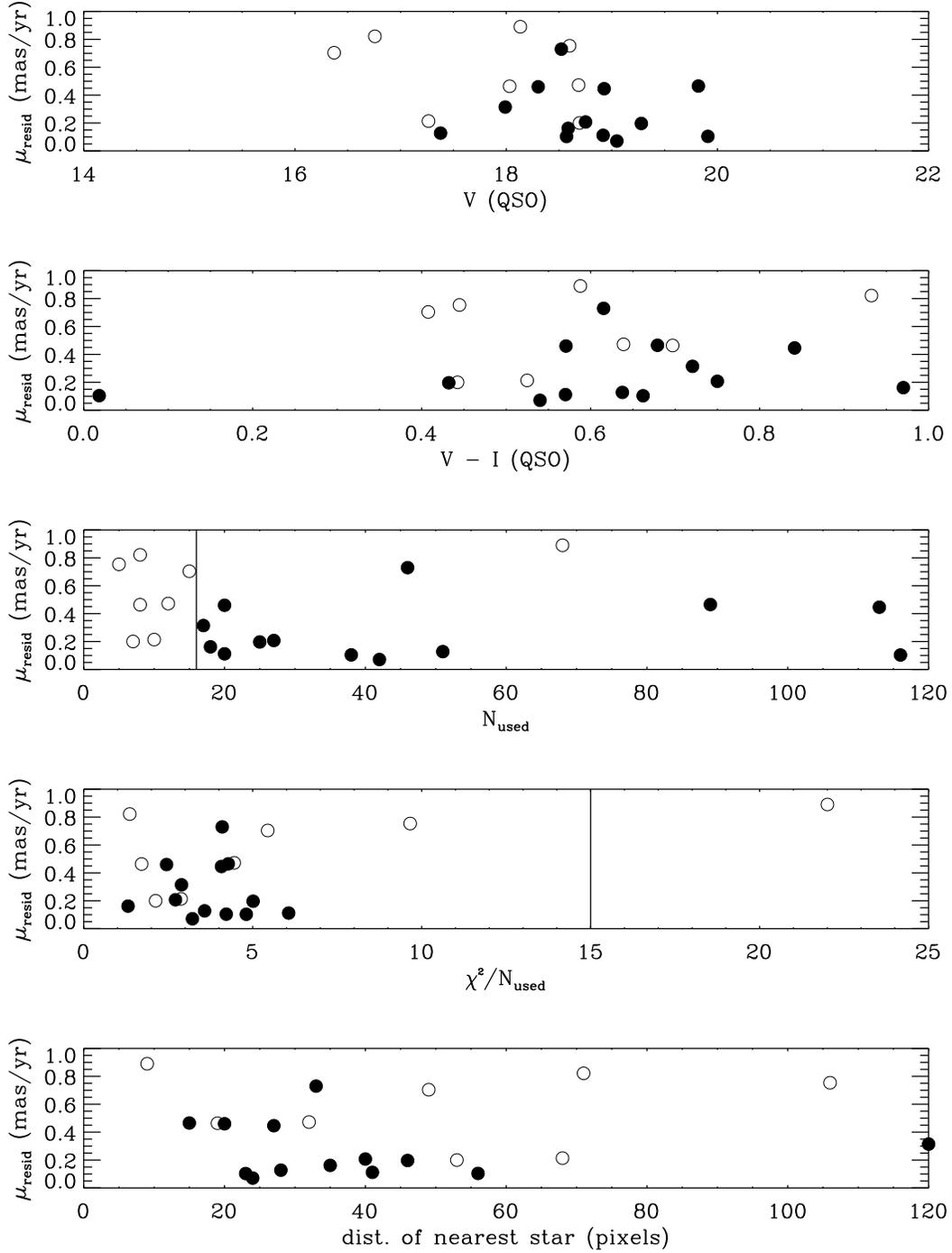}}
\caption{Plots of $\mu_{\rm {resid}}$ as a function of QSO $V$ magnitude, 
$V-I$ color, $N_{\rm {used}}$, $\chi^2/N_{\rm {used}}$ and distance to the nearest 
neighboring star. To select only the highest quality fields (closed circles), we 
retained only fields with 
$N_{\rm {used}} > 16$ and $\chi^2/N_{\rm {used}} < 15$ (vertical lines) in 
our final sample. The fields that are rejected on the basis of these cuts are shown with 
open circles in each panel.}
\label{finalcuts}
\end{figure}

The quantities $\PM_{\rm est}({\rm CM})$ for each field all provide an
independent estimate of the same quantity. For each field we
calculated the residual vector $\mu_{\rm{resid}}$ between its value of
$\PM_{\rm est}({\rm CM})$ and the final estimate of the LMC center of
mass proper motion that we derived from these estimates (to be
discussed in \S4.3 below). The scatter in these residuals
provides insight into the observational uncertainties. To address the
possibility of systematic trends we show in Figure~\ref{finalcuts} the size $|
\mu_{\rm{resid}} |$ of the residuals as a function of the following
quantities: the $V$ magnitude of the QSO; $V-I$ of the QSO; 
the number of stars $N_{\rm
{used}}$ used to align the stars in each field between the two epochs;
the quantity $\chi^2/N_{\rm {used}}$ that indicates how well the stars
could be aligned between the epochs; and the distance of the QSO to
its nearest neighboring star (without application of cuts to the star-list).

There is a clear systematic effect as a function of
$\chi^2/N_{\rm {used}}$. There is one field (ID L13) that has a much
higher value of $\chi^2/N_{\rm {used}}$ than the other fields. This
field also happens to have the largest residual $\mu_{\rm{resid}}$. It  
also has the nearest neighboring star of all fields (only 9 pixels 
from the QSO). So it is not clear whether this is the reason for the 
large  $\mu_{\rm{resid}}$ or whether it is the poor accuracy (measured 
by  $\chi^2/N_{\rm {used}}$) with which the star-field could be aligned. There
is also a systematic effect as a function of $N_{\rm used}$.  Fields
with a very low value of $N_{\rm {used}}$ do tend in general to have larger 
residuals than fields with a high value of $N_{\rm {used}}$. This makes sense,
since one expects to be able to align the data from the two epochs
better if there are more stars in the field, and we certainly do not 
want to trust the fields in which very few stars were used to determine the 
linear transformations. Thus, based on these plots we
decided to retain for our final estimate of the LMC proper motion only
the 13 fields for which $N_{\rm {used}} > 16$ and $\chi^2/N_{\rm {used}} <
15$. These fields are shown with filled circles in Figure~\ref{finalcuts} while  
the rejected fields are shown with open circles. Once we have applied these cuts 
there are no trends in $\mu_{\rm{resid}}$ for the remaining fields with 
 QSO magnitude, color, or distance to the nearest star. Hence, no additional 
cuts were applied.

\begin{figure}
\centerline{%
\epsfxsize=0.65\hsize
\epsfbox{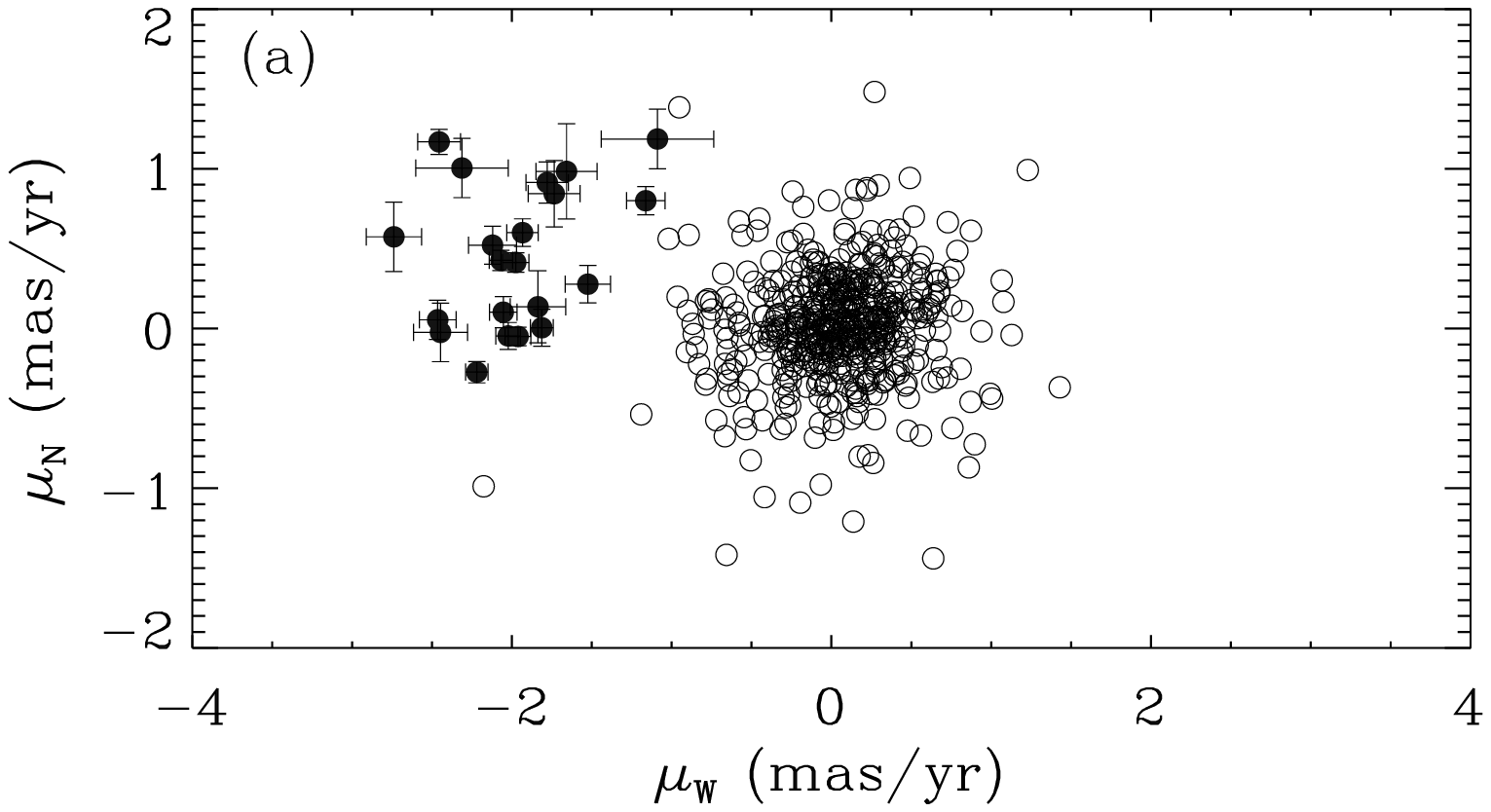}
\epsfxsize=0.65\hsize
\epsfbox{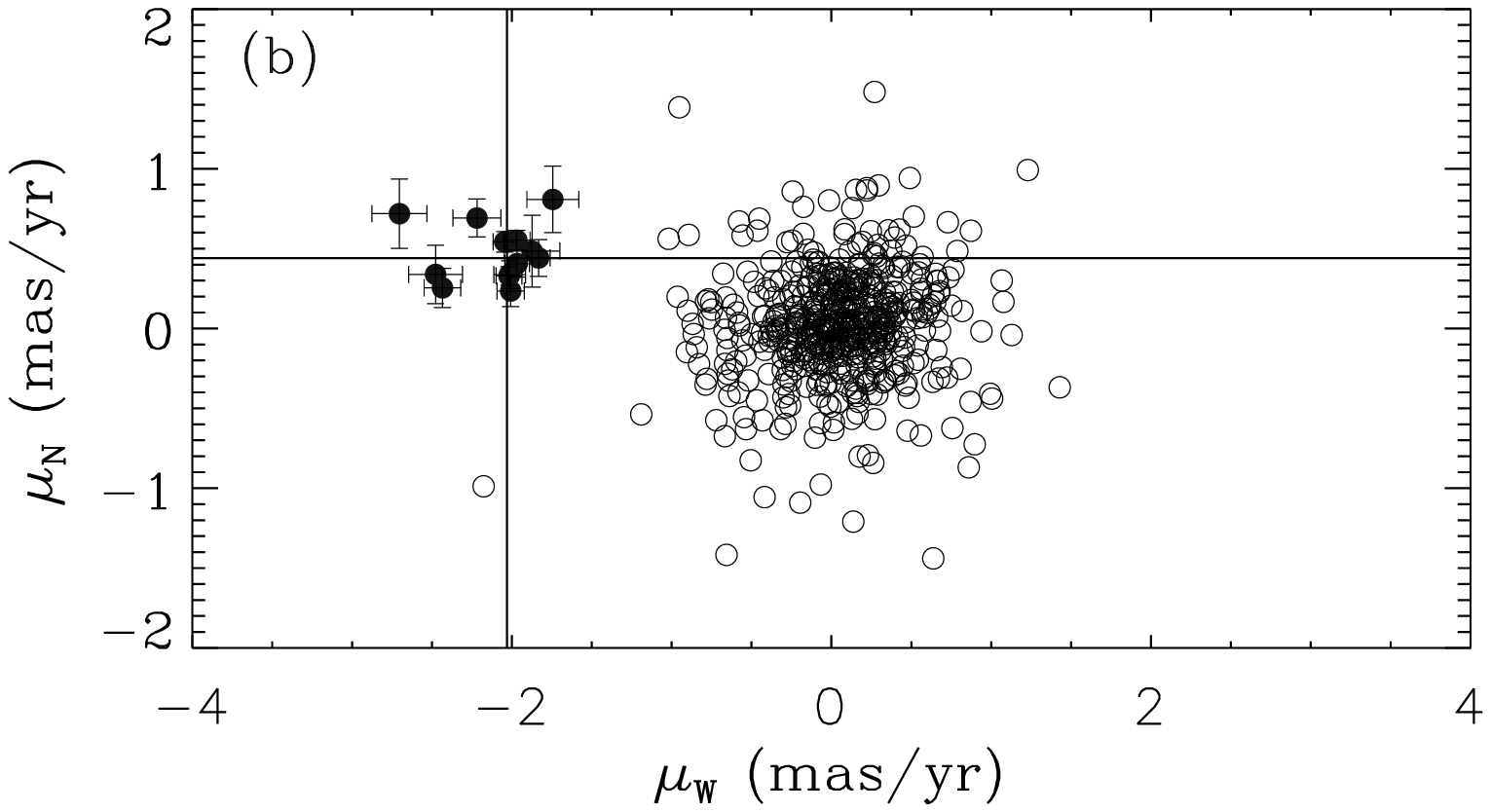}}
\caption{(a) The observed PM ($\mu_{W}$, $\mu_N$) for all the QSO fields (i.e., 
$-1 \  \times$ the observed reflex motion of the QSO; columns 5 \& 6 of Table~2); (b) 
the estimates $\PM_{\rm est}({\rm CM})$ of the LMC center of mass proper 
motion (columns 9 \& 10 
of Table~2) for the 13 ``high quality'' fields. These include corrections for 
viewing perspective and internal rotation. The residual PMs of the LMC 
stars in all the fields are plotted with open circles in both panels. The reflex motions 
of the QSOs clearly separate from the star motions. 
The straight lines in (b) mark the weighted average of the 13 fields, as listed in 
equation~(\ref{PMfinal}).}
\label{PMs}
\end{figure}

The PM values for all the QSO fields are shown in the ($\mu_W,\mu_N$)-plane in 
Figure~\ref{PMs}. They are compared to the residual proper motions of all the LMC 
stars that were used in the transformations. The stars are shown with 
open circles and the QSOs with filled ones. Panel (a) 
shows the observed PM values for all fields. Panel (b) shows the 
estimates $\PM_{\rm est}({\rm CM})$ derived from the observed values for 
the final 13 ``high-quality'' 
fields only. The reflex motion of the QSOs clearly separates from the star-fields 
in both panels. The solid lines
in (b) mark the weighted average of the 13 fields (see \S4.3, equation~(\ref{PMfinal})).

\begin{figure}
\plotone{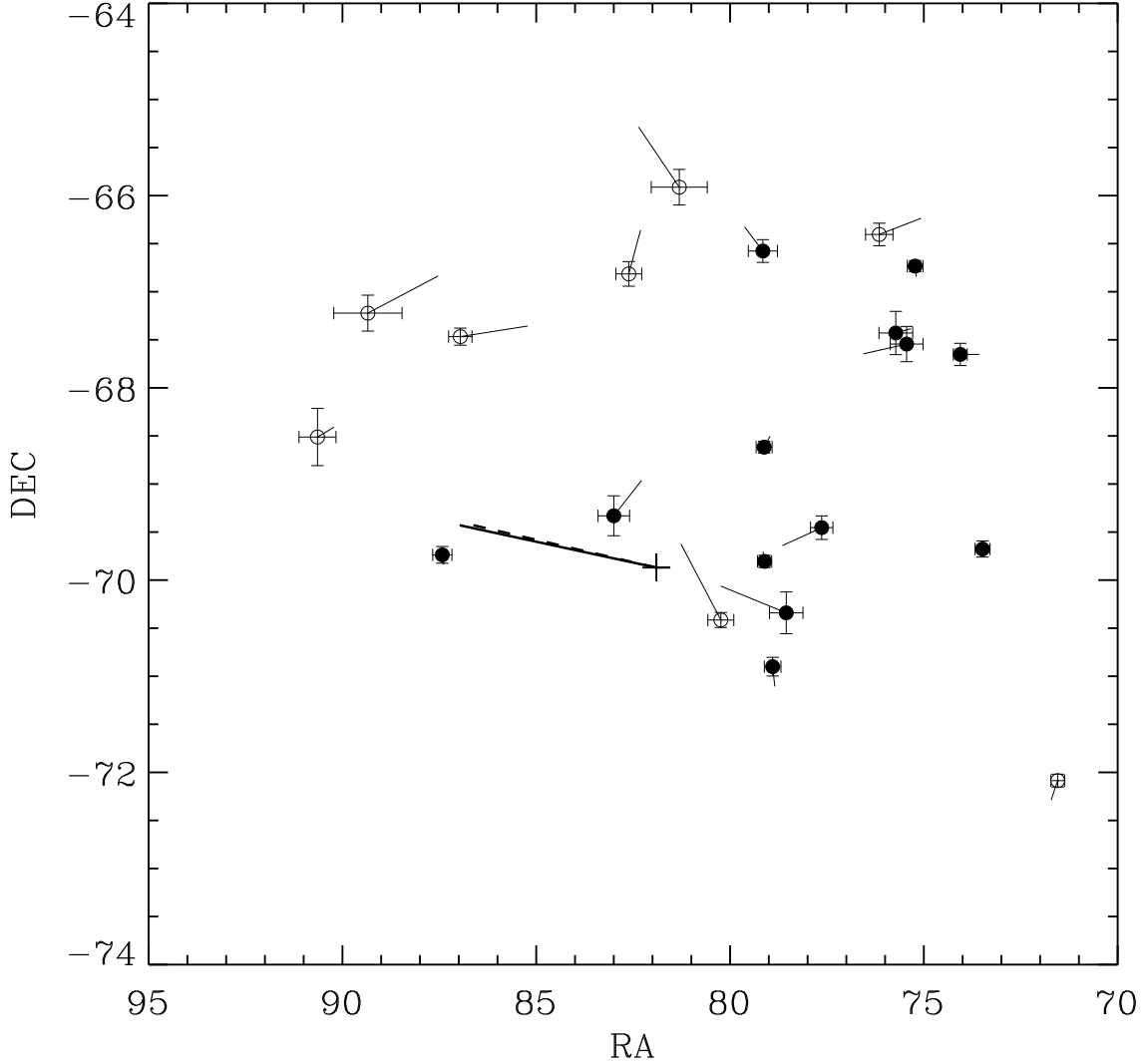}
\caption{Circles show the positions of the QSO fields. The 
fields that are rejected (from our final LMC PM estimate) 
are shown with open circles and the 
13 ``high quality'' fields are shown with filled circles. The error 
bars for each field are plotted as well. The vectors at these 
circles show the residuals between the PM estimates, $\PM_{\rm est}({\rm CM})$, derived 
from the data for these fields and the adopted weighted average. The latter 
is given in equation~(\ref{PMfinal}) and is shown by the bold solid vector that 
is anchored by a plus sign. The dashed vector (also anchored by the plus sign) 
represents the straight average of all 21 fields (columns 5 \& 6 of Table~2).} 
\label{mu_resid}
\end{figure}

Figure~\ref{mu_resid} shows the residual vectors $\mu_{\rm{resid}}$  
as a function of field position along with the $1-\sigma$ error bars for 
each field. 
Rejected and non-rejected fields are shown 
with open and filled circles respectively. The thick solid vector anchored 
by a plus sign shows for comparison the size of the inferred center of mass 
proper motion of the LMC, at the adopted 
LMC center. For the high quality fields there is no obvious trend 
as a function of absolute telescope pointing. Low quality fields 
that were rejected on the basis of 
their $N_{\rm{used}}$ or $\chi^2/N_{\rm{used}}$ do tend to have large 
residuals and appear to be concentrated mainly in the North-East. 
This  reflects 
the low stellar density in this region (see Figure~\ref{figure1}). 
On the other hand, the LMC is known to have internal motions and 
these  fields appear to both correspond to the location of the LMC 
spiral arm (Nikolaev \etal 2004) and show a similar 
residual trend in direction. By not including these fields we  
thus risk removing a real and interesting trend from the data. However,
 given the very low numbers of stars in these fields we do not feel 
that we can detect such a trend with confidence. Thus, these fields 
are not included in our final estimates. Note that if we were to 
include them, they would not make a  statistically significant 
difference to the bulk motion of the LMC (see \S4.3). This is 
illustrated by the bold dashed line in Figure~\ref{mu_resid} which 
is a straight average of all 21 fields (columns 5 \& 6 of Table~2).

\subsection{Influence of CTE degradation}

One possible source of systematic errors in our analysis is the
degrading Charge Transfer Efficiency (CTE) of the ACS/HRC CCD. When
charge is read out from a CCD pixel, some charge is left behind and
released at a later time. This causes sources in an image to show
faint tails along the detector $y$-direction. The impact of imperfect
CTE is larger for sources that are further from the read-out
amplifier. The impact also increases with time, because the continuing
bombardment by cosmic rays in the harsh environment of space
increases the number of charge traps in the CCD. The effect of
imperfect CTE on photometry has been studied in detail by many
authors, and has been quantified for the ACS/HRC by Riess (2003).
More recently, some attention has been drawn to the impact of
imperfect CTE on astrometry (Bristow, Piatek \& Pryor 2005; Piatek
\etal 2005). These authors quantified this effect for the STIS
instrument on $HST$ in the context of a study of the proper motion of
the Ursa Minor dwarf spheroidal galaxy. They found that the inferred
reflex PM of the QSO in their STIS field changes by $0.47\masyr$ due
to imperfect CTE.

The ACS/HRC CCD was a flight-spare for the STIS instrument, and its
detector properties are similar to those of the STIS CCD.  However,
the effect of imperfect CTE on our LMC PM should be much less than
that found in the STIS Ursa Minor study, for several reasons. First,
the ACS/HRC pixels span only $\sim 28$ mas on the sky, as compared to
$50$ mas for the STIS CCD. So the same star trail in pixel space will
produce an astrometric shift in mas that is smaller by a factor of $1.8$.
Second, the median observation date of our second epoch LMC data was
only 2.5 years after installation of the ACS on $HST$, as compared to 5.0
years for the last epoch in the STIS Ursa Minor study. If the effect
of imperfect CTE on astrometry varies linearly with time since
installation (as it does for photometry; Mutchler \& Sirianni 2005)
then our study should be affected by a factor of $2.0$ less. And third,
we have observations for $N=21$ different fields (or $N=13$ for the
``high quality'' sample), as compared to only 1 in the STIS Ursa Minor
study. Since our fields have more or less random detector orientations
on the sky, any CTE induced astrometric shift will average down to
zero $\propto N^{-1/2}$. So even if we were to make no allowance for
CTE one would not expect our study to be affected by more than $\sim
0.04 \masyr$.

The STIS Ursa Minor study could correct explictly for the astrometric
effects of CTE owing to the existence of a detailed model for the
underlying STIS CCD detector physics and charge trap properties
(Bristow \& Alexov 2002). Such a model does not (yet) exist for the
ACS/HRC CCD, and it is therefore not possible to correct our data
explictly for imperfect CTE. However, a simple correction can be
applied by fitting a shift in the detector $y$-direction of sources
that is linearly proportional to both the $y$-position on the detector
and the time since installation of the instrument in space. This is
akin to what Piatek \etal (2005) did for the WFPC2 data that they
obtained for Ursa Minor. They had to include this term explictly,
because they otherwise allowed only for a translation, rotation and
scale change between epochs (i.e., a four-parameter linear
fit). Instead, in our study we perform a fully general six-parameter
linear fit to align the star-fields from different epochs. Therefore,
the type of astrometric shift expected from imperfect CTE is
automatically allowed for and fitted to the data. 
This seems to work well, given that the average PMs of the LMC stars
  do not show a trend with $y$-position on the detector (see 
Figure~\ref{pmbins}). The only thing that
we do not allow for is a variation of the astrometric shift with
brightness. Brighter sources are less affected by imperfect CTE than
fainter sources. Since the QSO is generally brighter than the average
LMC star in the field, there might be a small residual CTE effect in
the PM for each field. However, we will estimate the final errors in
the LMC PM from the scatter between the PM measurements for different
fields. Therefore, any residual systematic errors due to imperfect CTE
are explictly accounted for in our final error estimates.

So in summary: (1) even when not modeled at all, imperfect CTE is
expected to have at most a small effect on our ACS/HRC results for the
LMC PM; (2) our methodology for aligning star fields between epochs
allows for a low-order correction of the astrometric shifts due to
imperfect CTE; and (3) any residual effects of imperfect CTE are
explictly accounted for in our final error estimates.

\subsection{Proper Motion of the LMC center of Mass}

To obtain our final estimate of the LMC proper motion we took the
weighted average of the 13 high-quality fields to obtain
\begin{equation}
\label{PMfinal}
   \mu_W = -2.03 \pm 0.08 \ \rm{mas \ yr^{-1}}, \;
   \mu_N =  0.44 \pm 0.05 \ {\rm mas \ yr^{-1}} \  (HST) . 
\end{equation}
The errors in these numbers are based on the RMS scatter between
different fields, divided by $\sqrt{N}$. To be conservative we did not
use the weighted average errors $[ \sum \{ \delta \PM({\rm field})
\}^{-2} ]^{-1/2}$. These errors have a value of $0.03\masyr$ to the North
and to the West, and are a factor of $\sim 2$ smaller in the Northward 
direction and $\sim3$ smaller in the Westward direction 
than the errors in equation~(\ref{PMfinal}). The fact that the
weighted errors, which are due to pure noise propagation, are smaller
than the errors in equation~(\ref{PMfinal}) indicates that there is more scatter in the
results from different fields than can be accounted for by random
errors alone.  This is not entirely unreasonable. An error of
$0.07\masyr$ over a baseline of 1.6 years (our mean) corresponds to
0.004 pixels, and it has not been established that there are no
systematic errors in, for example, the geometric distortion, at that
level (see Figure~\ref{pmbins} and AK04). One possible cause of systematic 
errors might be variations in
the higher order geometric distortion terms with time at the level of
thousandths of a pixel or residual
astrometric effects due to imperfect CTE. Note that if so, such systematic errors would
probably have a fixed orientation with respect to the detector.  Since
our fields were observed at somewhat random orientations because of
the snapshot nature of our program (see Figure~\ref{figure2}), such 
systematic errors 
would still approximately decrease as $1/\sqrt{N}$ when averaging 
(as we assumed in getting equation~(\ref{PMfinal})).

It is important to note that our result does not depend sensitively on
the details of either the modeling of the field-dependence of the PMs, or
the criteria applied to reject lower-quality fields, or the actual
statistic applied to obtain the average. A straight unweighted average
of the observed proper motions for all 21 fields, as listed in
Table~2 (columns 5 \& 6), yields $\mu_W = -1.97 \pm 0.09 \masyr, \;
\mu_N = 0.46 \pm 0.10 \masyr$.
The median is $(-1.98, \ 0.43) \masyr$. These estimates 
are both consistent with the result in equation~(\ref{PMfinal}) to
within the errors.

\subsection{Proper Motion Rotation of the LMC}     

\begin{figure}
\centerline{%
\epsfxsize=0.65\hsize
\epsfbox{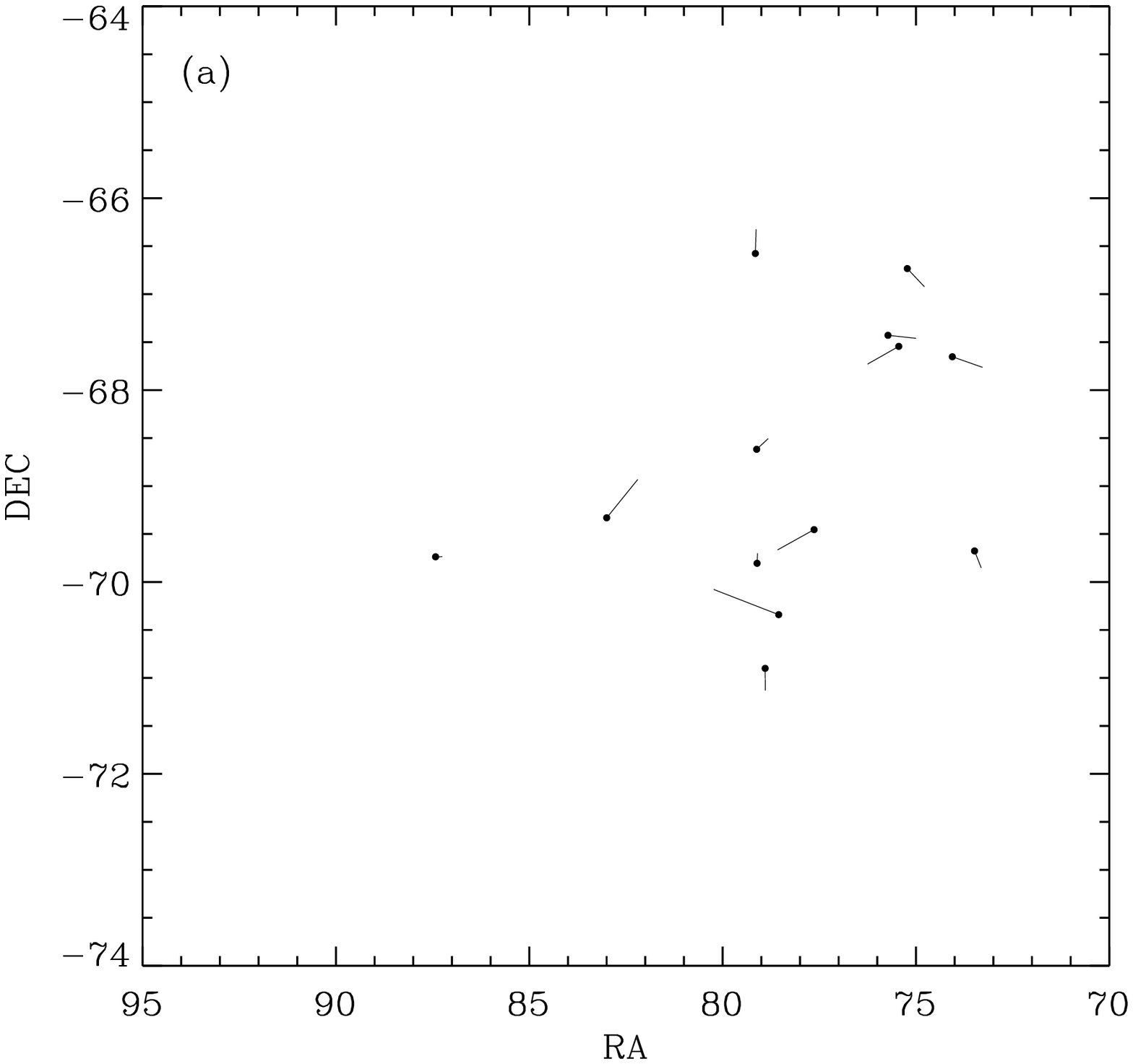}
\epsfxsize=0.58\hsize
\epsfbox{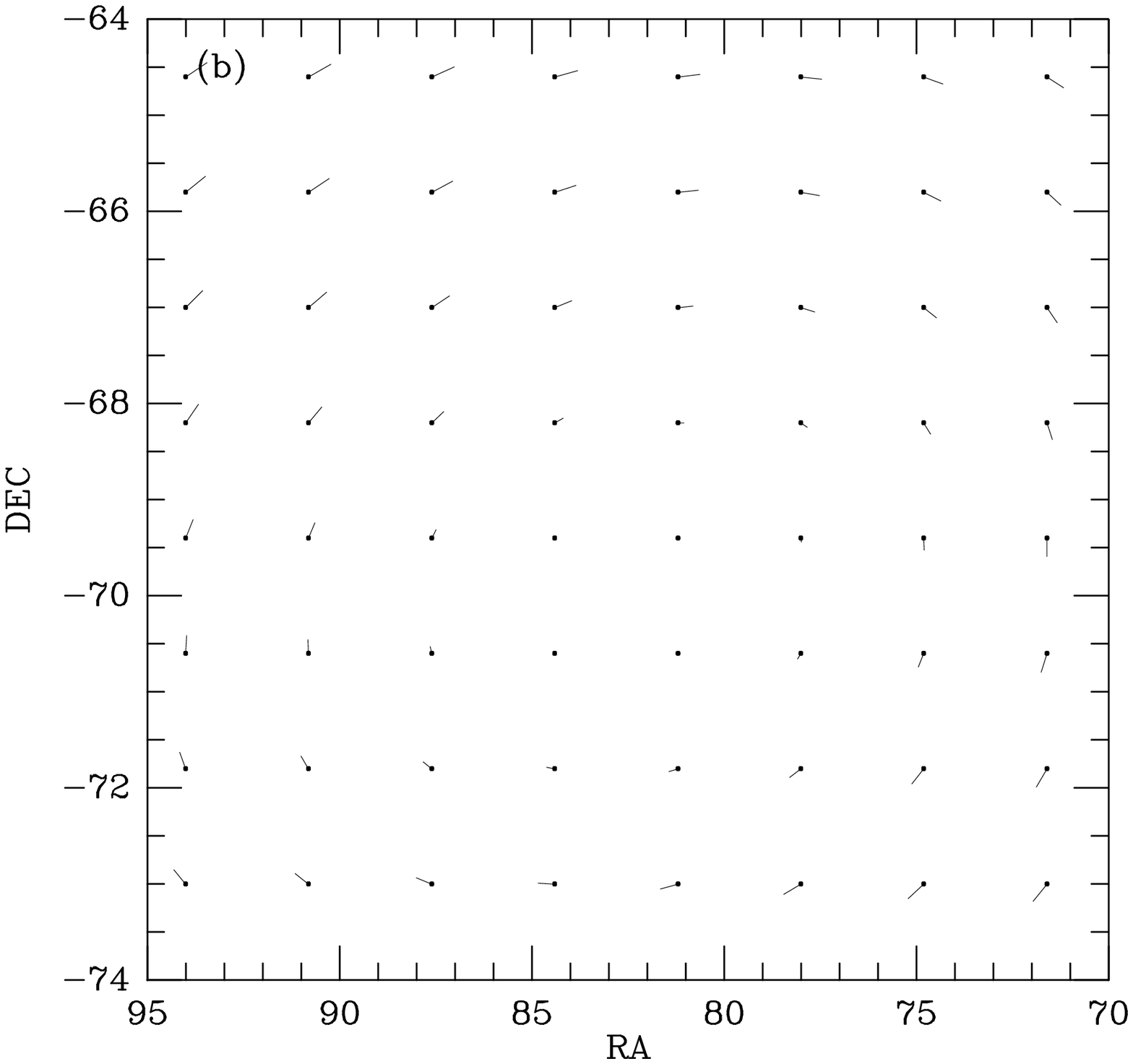}}
\caption{(a) LMC internal proper motion rotation estimates 
$\hat \mu_{\rm{resid}}$ for the high quality fields. 
(b) The expected rotation field of the 
LMC from the best fit model for the line-of-sight velocities 
of carbon stars described in vdM02 and \S 4.1.}
\label{rotation}
\end{figure}

Our data might in principle allow us to detect the rotation of the LMC
in the plane of the sky. To address this we plot in 
Figure~\ref{rotation}(a) 
the residuals  
\begin{equation}
  \hat{\mu} = 
    \PM({\rm field}) - \PM({\rm CM}) - \PM_{\rm res}({\rm field}) ,
\end{equation}
where $\PM({\rm field})$ is the observed proper motion for a field and 
$\PM({\rm CM})$ is the center-of-mass proper motion from
equation~(\ref{PMfinal}). The quantity $\PM_{\rm res}({\rm field})$ is
the residual proper motion as in \S4.1, but now includes only
the contribution from viewing perspective, and {\it not} the
contribution from internal rotation. The residuals ${\hat \mu}$ are
therefore the observational estimates of the internal motions of the
LMC. For comparison, we show in Figure~\ref{rotation}(b) 
the internal motions predicted by the model employed in \S4.1.

It is clear that the observations are dominated by noise, and 
rotation is not immediately obvious from visual inspection. On the other
hand, some signal can be measured by averaging the results
from the individual fields. For each field we determined the unit
vectors ${\vec u}_{\parallel}$ and ${\vec u}_{\perp}$ along and
perpendicular to the direction of rotation predicted by the model.
 We determined the components of the residual
vectors ${\vec{\hat \mu}}$ along these unit vectors, which are given by
the inner products $\hat \mu_{\parallel} = {\vec {\hat \mu}} \cdot {\vec
u}_{\parallel}$ and $\hat{ \mu_{\perp}} = {\vec {\hat \mu}} \cdot {\vec
u}_{\perp}$. We then determined the averages of these
components for our 13 high-quality fields. This yields $\langle
\hat {\mu_{\parallel}} \rangle = 0.09 \pm 0.07 \masyr$ and $\langle
\hat {\mu_{\perp}} \rangle = -0.06 \pm 0.07 \masyr$. For comparison, the
model predicts $\langle \hat {\mu_{\parallel}} \rangle = 0.08 \masyr$ and
$\langle \hat{ \mu_{\perp}} \rangle = 0$. Although the errors are
significant, we do detect the expected sign and the expected
magnitude for the rotation. This provides additional confidence in the
accuracy of our results.

In our models we have assumed throughout that $di/dt = d\Theta/dt =
0$. If these terms were to have non-zero values then this would have
left a signature in the residuals ${\vec {\hat \mu}}$. In particular,
using model calculations similar to those in Figure~10c,d of vdM02,
and assuming that the model for the internal rotation of the LMC is
correct, we would have expected that in $\masyr$, $\langle \hat {\mu_{\parallel}}
\rangle = 0.08 - 0.046 \ d\Theta/dt$ and $\langle \hat \mu_{\perp} \rangle
= -0.01 \ di/dt$. In these equations both $d\Theta/dt$ and $di/dt$ are also in
units of $\masyr$ (where $1 \masyr = 278^{\circ}/{\rm Gyr}$). The
observed values therefore provide the following constraints on the
precession and nutation of the LMC:
\begin{equation} 
  d\Theta/dt = -0.17 \pm 1.57 \masyr , \; 
  di/dt = -6 \pm 7 \masyr .
\end{equation}
Both quantities are consistent with zero,
  although the constraints are not particularly informative. The error
  bars are considerably larger than the precession and nutation that are
expected on the basis of $N$-body calculations (Weinberg 2000).

\section{Discussion}

\subsection{Comparison to Previous Proper Motion Work}
\begin{figure}
\plotone{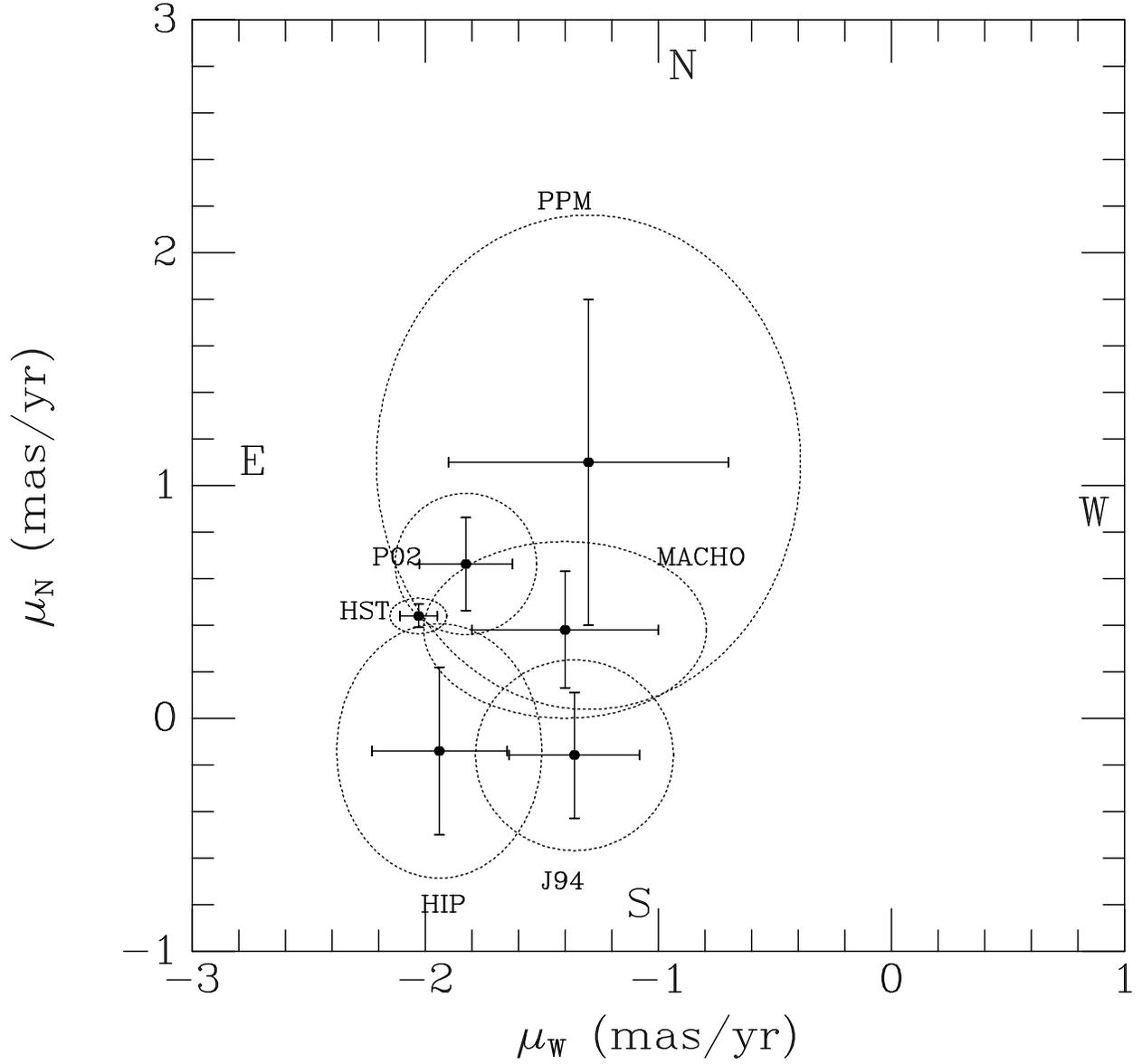}
\caption{The ($\mu_W,\mu_N$)-plane spanned by the proper motion 
components of the LMC center of mass from various studies. Dotted 
ellipses are the corresponding $68.3$\% confidence regions. PPM 
stands for the Kroupa \etal (1994) study, P02 for Pedreros \etal 
(2002), MACHO for Drake \etal (2001), HIP for Kroupa \& Bastian (1997), 
J94 for Jones \etal (1994), and HST for this study.}
\label{compPM}
\end{figure}

We are now in a position to make comparisons with previous proper motion 
work. Figure~\ref{compPM} shows all the PM determinations to date with 
their $68.3$\% confidence ellipses. We do not include the very discrepant 
results of Anguita \etal 
(2000) and Momany \& Zaggia (2005) on this plot as there must be unidentified 
systematic errors associated with these measurements (see discussions in Pedreros 
\etal 2000 and Momany \& Zaggia 2005). The figure shows 
the region of the $(\mu_W, \mu_N)$-plane spanned by the rest of the 
results. These results are more-or-less consistent with each other.
 Our $HST$ 
determination appears to be closest to the results of the Pedreros \etal (2002)
study  and 
the Kroupa \& Bastian (1997) Hipparcos study. Our value for $\mu_W$ is on the 
high end of what was found in previous studies.

The measurements from all these studies can be combined to get a 
weighted average of the 
proper motion of the LMC. In doing so we first increase (as in vdM02) 
the error bars of all the studies by a fixed factor so that they become 
statistically consistent, as judged by $\chi^2$. The resulting grand 
average of all the LMC proper motion measurements is 
\begin{equation}
 \mu_W = -1.94 \pm 
0.09 \masyr, \ \mu_N = 0.43 \pm 0.06 \masyr \ (HST + \rm{other \ studies}).
\end{equation}
In the discussion that follows we adopt the $HST$-only values given in 
equation~(\ref{PMfinal}).

\subsection{Three-dimensional Space Motion of the LMC}

The observed proper motion of the LMC quoted in equation~(\ref{PMfinal}) 
yields an estimate of its transverse velocity via 
\begin{equation}
v_x = D_0\mu_W, \; v_y = D_0\mu_N,
\end{equation}
where the $x$- and $y$-directions point towards the direction of West
and North respectively. $D_0$ is the LMC distance, which we assume to
be $50.1$ kpc as in \S4.1 (Freedman \etal 2001). This yields
\begin{equation} 
  v_x = -482 \pm 18\kms,\; v_y = 104 \pm 12\kms.
\end{equation}
The listed errors do not include the potential contribution from the 
uncertainty in the distance modulus. The values of $v_x$ and $v_y$ 
correspond to a transverse velocity of
$v_t=493\kms$ in the direction of position angle $\Theta_t=78\degr$.
The systemic line-of-sight velocity is very well known from a number
of different tracers including a detailed analysis of 1041 carbon
stars by vdM02 who obtained $v_{sys} =
262.2\pm3.4\kms$.  We are therefore in a position to ask what these
measurements imply for the three-dimensional space motion of the
LMC. This issue has been addressed previously by several authors but
our results provide the most accurate measurement so far.

As many previous authors have done we adopt a Cartesian coordinate system 
($X$, $Y$, $Z$) with the origin at the Galactic center, with the $Z$-axis 
pointing toward the Galactic north pole, the $X$-axis pointing in the 
direction from the Sun to the Galactic center, and the $Y$-axis 
pointing in the direction of the Sun's Galactic rotation. It is 
necessary to correct for the reflex motion of the Sun to determine the 
motion of the LMC with respect to the Milky Way. Following the equations 
set up in vdM02,  \S9.3, we get
\begin{eqnarray}
\mathbf{v}_{\rm{LMC}} = (-86\pm12,-268\pm11,252\pm16)\kms,\nonumber\\
v_{\rm{LMC}} = 378\pm18\kms,\nonumber\\
v_{\rm{LMC,rad}} = 89 \pm 4\kms, \  v_{\rm{LMC,tan}} = 367 \pm 18\kms,
\end{eqnarray}
for the 3-D velocity of the LMC and its radial and tangential 
components. The fact that $v_{\rm{LMC,tan}}$ is larger than the 
Milky Way circular velocity of $\sim220\kms$ and that 
$v_{\rm{LMC,rad}}$ is small and positive confirms that the LMC 
is just past perigalacticon, as found by previous authors.

\subsection{Comparison to Magellanic Stream Models}
 Our value for the galactocentric tangential velocity is higher 
than that implied by a weighted average of earlier PM measurements, 
 which is $281\kms$. It is also a little higher than 
values inferred from most Magellanic Stream models. There are two main approaches 
to modeling the Stream, one that invokes ram-pressure stripping of MC gas 
by matter in the Galactic Halo, and another that invokes the tidal 
force of the Milky Way on the MC system. As stated in the introduction, 
 Gardiner \& Noguchi (1996) have constructed detailed simulations for the 
tidal distortion of the SMC owing to the Galaxy and the LMC, and get 
the following constraints for the motion of the LMC: 
$\mathbf{v}_{\rm{LMC}} = (-5,-225,194)\kms, \  
v_{\rm{LMC,rad}} = 80\kms,$ and $v_{\rm{LMC,tan}} = 286\kms$. Our observed 
value for the tangential velocity is noticably higher, while the radial 
velocity seems to agree well.
Our values appear more consistent with those of Heller \& Rohlfs (1994) 
who use a combination of the tidal and ram-pressure stripping models to  
obtain : $\mathbf{v}_{\rm{LMC}} = (-10,-287,230)\kms, \  
v_{\rm{LMC,rad}} = 107\kms,$ and $v_{\rm{LMC,tan}} = 352\kms$. 
The errors in the proper motion of the LMC are now technically 
good enough to allow distinctions between various Magellanic Stream models. Our 
$HST$  proper 
motion determination for the SMC will soon follow in a subsequent paper. 
Taken together 
these two data points should provide  powerful new constraints on the MC 
system and the gravitational potential of the Galactic Halo.

\subsection{Other New Insights}  

Our results also lead to revised values for several other quantities that 
characterize the LMC. Given a revised PM we can obtain a different 
rotation curve $V(R)$ for the carbon stars in the LMC. 
This results in an amplitude of 
$60 \kms$ for the LMC rotation 
curve which is an increase of $\sim20$\% compared to the vdM02 value. This will 
in turn cause small changes in other estimates that are based on $V(R)$ such
as the total mass of the LMC and its tidal radius.

Updated modeling of the carbon star velocity field using our revised 
PM values also yields an 
improved constraint on $di/dt$. Using the same methodology as in vdM02 
we obtain $di/dt = -0.12\pm0.09\masyr$, 
which is now more consistent with zero than previously believed. This is a 
much stronger constraint than what we were able to obtain with the analysis of the 
field-dependent PMs alone (equation~(5)).

\section{Summary \& Conclusions}

We undertook a project using two epochs of $HST$/ACS data of Magellanic Cloud 
fields centered on background QSOs to determine the systemic proper motion 
of the Clouds. The LMC results are presented in this paper. We have determined 
the proper motion of the LMC to be 
$ \mu_W = -2.03 \pm 0.08 \ \rm{mas \ yr^{-1}}, \;
	  \mu_N = 0.44 \pm 0.05 \ \rm{mas \ yr^{-1}}$. 
This is accurate to better than 5\%. When combined with HI data for the 
Magellanic Stream, this should allow improved constraints on both the mass distribution in 
the Galactic Halo, and theoretical models for the origin of the Magellanic Stream. 
 Our data provides the most accurate proper motion measurement for any Milky Way satellite.
In future papers, we will present results for the SMC as well as 
implications for the LMC-SMC system and the Magellanic Stream.

Improvements to our work may be possible  by using $HST$ with 
a longer baseline. A long baseline would facilitate measurements of the 
internal motions of the Clouds as well as an even more accurate 
measurement of their systemic motions. In addition we may be able to make  
a distance determination using the method of rotational parallax 
(Olling \& Peterson, 2000).  
 Rotational parallax is the method of determining distances
to local group galaxies by measuring a rotation curve using both
proper motions and radial velocities (see Brunthaler \etal 2005 for an
application of this method to M33). Since the former are
distance-dependent and the latter are not, using both methods provides
a distance measurement. More generally, considerable improvement may be 
expected when the next generation of astrometric satellites such as $SIM$ 
and $GAIA$ come on-line.

\acknowledgments
The authors would like to thank Jay Anderson and Ivan King for their 
geometric distortion calibration of the HRC, and in particular, Jay Anderson 
for making his analysis software available to the community.
Support for this work was 
provided by NASA through grant numbers
GO-09462 and GO-10130  from the SPACE TELESCOPE SCIENCE INSTITUTE (STScI),
which is operated by the Association of Universities for Research in
Astronomy, Inc., under NASA contract NAS5-26555.  M.~G.~is supported by NASA through Hubble
 Fellowship grant HF-01159.01-A awarded by STScI. KHC's work was performed under the 
auspices of the U.S. Department of Energy, National Nuclear Security 
Administration by the University of California, Lawrence Livermore National 
Laboratory under contract No. W-7405-Eng-48.

\section{References}
\noindent Alcock, C. 2000, Science, 287, 74\\
\noindent Alcock, C. \etal 2000a, ApJ, 542, 281\\
\noindent Alcock, C., et al.\ 
2000b, \aj, 119, 2194\\ 
\noindent Anderson, J. \& King, I.~R.\ 2004, ACS Instrument Science Report 04-15 
(Baltimore: Space Telescope Science Institute) (AK04)\\
\noindent Anderson, J., \& 
King, I.~R.\ 2003, \pasp, 115, 113\\
\noindent Anguita, C., Loyola, P., \& Pedreros, M.~H. 2000, AJ, 120, 845\\
\noindent Bristow, P., \& Alexov, A. 2002, ST-ECF Instrument Science Report
  CE-STIS 2002-01 (Garching bei Munchen: Space Telescope European
  Coordinating Facility)\\
\noindent Bristow, P., Piatek, S., \& Pryor, C. 2005, in ST-ECF Newsletter 38, p. 12
  (Garching bei Munchen: Space Telescope European Coordinating Facility)\\
\noindent Br{\" u}ns, C., et al.\ 2005, \aap, 432, 45\\ 
\noindent Brunthaler, A., 
Reid, M.~J., Falcke, H., Greenhill, L.~J., \& Henkel, C.\ 2005, Science, 
307, 1440 \\
\noindent Drake, A.~J., Cook, 
K.~H., Alcock, C., Axelrod, T.~S., Geha, M., \& MACHO Collaboration 2001, 
BAAS, 33, 1379\\
\noindent Dobrzycki, A., Macri, 
L.~M., Stanek, K.~Z., \& Groot, P.~J.\ 2003, \aj, 125, 1330\\
\noindent Dobrzycki, A., Groot, 
P.~J., Macri, L.~M., \& Stanek, K.~Z.\ 2002, \apjl, 569, L15\\ 
\noindent Fich, M.~\& Tremaine, S. 1991, ARA\&A, 29, 409\\
\noindent Freedman, W.~L., et 
al.\ 2001, \apj, 553, 47\\
\noindent Gardiner, L.~T., 
\& Noguchi, M.\ 1996, \mnras, 278, 191 \\
\noindent Geha, M.~et al. 2003, AJ, 125, 1\\
\noindent Heller, P.~\& Rohlfs, K. 1994, A\&A, 291, 743\\
\noindent Jones, B.~F., Klemola, A.~R., \& Lin, D.~N.~C. 1994, AJ, 107, 1333\\
\noindent Karakla, D.~ 2005, $HST$ Primer (Baltimore: Space Telescope Science Institute)\\ 
\noindent Krist, J.\ 2003, ACS Instrument Science Report 03-06 (Baltimore: Space 
Telescope Science Institute)\\
\noindent Kroupa, P., R{\" o}ser, S., \& Bastian, U. 1994, MNRAS, 266, 412\\
\noindent Kroupa, P.~\& Bastian, U. 1997, New Astronomy, 2, 77\\
\noindent Lin, D.~N.~C., Jones, B.~F., \& Klemola, A.~R. 1995, ApJ, 439, 652\\
\noindent Merrifield, M. 1992, AJ, 103, 1552\\
\noindent Momany, Y., \& Zaggia, S.\ 2005, \aap, 437, 339\\ 
\noindent Mutchler, M., \& Sirianni, M. 2005, ACS Instrument Science Report
  2005-03 (Baltimore: Space Telescope Science Institute)\\
\noindent Nikolaev, S., Drake, A.~J., Keller, S.~C., Cook, K.~H., 
Dalal, N., Griest, K., Welch, D.~L., \& 
Kanbur, S.~M.\ 2004, ApJ, 601, 260 \\
\noindent Olling, R.~P.~ \& Peterson, D.~M., 2000, astro-ph/0005484\\
\noindent Pavlovsky, C.\ 2005, ACS Instrument Handbook (Baltimore: Space Telescope Science 
Institute)\\
\noindent Pedreros, M. H., Anguita, C., \& Maza, J. 2002, AJ, 123, 1971\\
\noindent Smecker-Hane, 
T.~A., Cole, A.~A., Gallagher, J.~S., \& Stetson, P.~B.\ 2002, \apj, 566, 
239\\
\noindent Piatek, S., Pryor, C., Bristow, P., Olszewski, E. W., Harris, H. C.,
  Mateo, M., Minniti, D., \& Tinney, C. G. 2005, AJ, 130, 95\\
\noindent Riess, A. 2003, ACS Instrument Science Report 2003-09 (Baltimore:
  Space Telescope Science Institute)\\
\noindent van der Marel, R.~P.~2002, in `The Shapes of Galaxies and their
Halos', Natarajan P., ed., 202, 2002 (Singapore: World Scientific)\\
\noindent van der Marel, R.~P., Alves, D.~R., Hardy, E., \& Suntzeff, N.~B.\
2002, AJ, 124, 2639 (vdM02)\\
\noindent Weinberg, M.~D.\ 2000, \apj, 532, 922\\ 
\noindent Wilkinson, M.I., \& Evans, N.W. 1999, MNRAS, 310, 645\\
\noindent Sofue, Y., \& Rubin, V.\ 2001, \araa, 39, 137 \\

\begin{deluxetable}{llllcccccccccccc}
\tabletypesize{\tiny}
\rotate
\tablewidth{0pt}
\tablecolumns{16}
\tablecaption{Sample and Observations}
\tablehead{
\colhead{ID}  & \colhead{QSOname} & 
\colhead{RA} & \colhead{DEC} &
\colhead{$V$} & \colhead{$z$} & \multicolumn{5}{c}{epoch1} 
& \multicolumn{4}{c}{epoch2} & \colhead{$\Delta time$}}
\startdata
 &  &  &  &  &  & date & visit 
& $T_{exp}$ & $T_{exp}$ & 
ORIENTAT & 
date & visit & $T_{exp}$ & ORIENTAT &   \\
  &   &  &  &  &  &  &  
& F606W & F814W &  &  &  &   F606W &  &   \\
 & & (H,M,S) & (deg, ', '')     &  &  &  &  & 
(min) & (min) & (deg) &  &  & (min) & (deg) & (yrs)\\
L1 & 68.10972.36 & 5 47 50.2 &  -67 28 1.3 & 16.4 & 1.01 & 2002-09-05
& 1 & 6.7 & 1.7 & 75.3 & 2005-04-25 & 1 & 6.7 & -51.3 & 2.6\\
L2 & 42.860.123 &   4 46 11.1  &    -72 5 9.0	& 17.3 & 0.26 &
2002-09-12 & 2 & 6.7 & 1.7 & 96.4 & 2004-09-14 & 2 & 6.7 & 96.5 & 2.0\\
L3 & 2.5873.82 &   5 16 28.9  &    -68 37 1.8 &	17.4 & 0.46 &
2002-09-24 & 3 & 6.7 & 1.7 & 100.9 & 2004-08-11 & 3 & 6.7 & 59.6 & 1.9\\
L4 & 58.5903.69  &  5 16 36.8   &   -66 34 35.8 &  18.0 & 2.24 &
2002-09-22 & 4 & 6.7 & 1.7 & 99.6 & 2004-09-12 & 4 & 6.7 & 90.3 & 2.0\\
L5 & 13.5717.178 &   5 15 36.1  &    -70 54 0.8 &	18.8 & 1.66 &
2002-09-07 & 5 & 6.7 & 1.7 & 68.8 & 2004-08-10 & 5 & 10.1 & 59.0 & 1.9\\
L6 & 75.13376.66 & 6 2 34.3 & -68 30 41.1 & 18.7 & 1.07 & 2002-10-11 &
6 & 6.7 & 1.7 & 106.9 & 2004-08-13 & 6 & 7.3 & 57.1 & 1.8\\
L7 & 78.5855.788  &  5 16 26.3   &   -69 48 19.0 &	18.6 & 0.63 &
2002-10-28 & 7 & 6.7 & 1.7 & 135.2 & 2004-08-10 & 7 & 7.3 & 58.7 & 1.8\\
L8 & 53.3360.344 &   5 0 54.0   &   -66 43 59.8   & 18.9 & 1.86 & 
2002-09-03 & 9 & 6.9 & 1.7 & 90.4 & 2004-09-05 & 9 & 11.5 & 90.5 & 2.0\\
L9 & 17.2227.488  &  4 53 56.5   &   -69 40 35.4  &  19.0 & 0.28 &
2002-09-05 & 11 & 9.5 & 1.7 & 88.6 & 2004-09-09 & 11 & 10.1 & 92.3 & 2.0\\
L10 & 30.11301.499  &  5 49 41.6  &    -69 44 15.1 &  19.9 & 0.46 &
2002-09-03 & 12 & 12.0 & 1.7 & 73.5 & 2004-08-10 & 12 & 21.5 &  48.9 & 1.9\\
L11 & 69.12549.21 & 5 57 22.4 & -67 13 21.5 & 16.8 & 0.14 & 2003-06-04 &
19 & 6.7 & 1.7 & -15.9 & 2005-04-28 & 18 & 6.7 & -51.4 & 1.9\\
L12 & 25.3469.117 &   5 1 46.7   &   -67 32 39.8 &  18.3 & 0.38 &
2003-06-04 & 20 & 6.7 & 1.7 & -3.4 & 2004-08-18 & 19 & 6.7 & 69.1 & 1.2\\
L13 & 6.6572.268  &  5 20 57.0  &    -70 24 52.6 &  18.1 & 1.81 &
2003-02-16 & 21 & 6.7 & 1.7 & -113.5 & 2004-08-10 & 20 & 6.7 & 57.8 & 1.5\\
L14 & 25.3712.72  &  5 2 53.7   &   -67 25 45.0 &  18.6 & 2.17 &
2003-06-07 & 24 & 6.7 & 1.7 & 12.1 & 2004-07-13 & 23 & 6.9 & 36.1 & 1.1\\
L15 & 9.5484.258  &  5 14 12.1   &   -70 20 25.8 &  18.5 & 2.32 &
2003-06-04 & 25 & 6.7 & 1.7 & -5.9 & 2004-07-14 & 24 &  8.0 & 33.3 & 1.1\\
L16 & 53.3970.140 & 5 4 36.0 & -66 24 15.7 & 18.0 & 2.04 & 2002-12-26
 & 26 & 6.7 & 1.7 &  -162.3 & 2005-04-07 & 25 & 6.7 & -59.5 & 2.3\\
L17 & 63.7365.151 & 5 25 14.4 & -65 54 45.7 & 18.6 & 0.65 & 2003-05-08 &
27 & 6.7 & 1.7 & -34.9 & 2004-07-05 & 26 & 8.8 & 30.3 & 1.2\\
L18 & 61.8199.302  &  5 30 26.8    &  -66 48 52.9 &  18.7 & 1.79 &
2002-12-25 & 29 & 6.7 & 1.7 & -169.3 & 2004-08-10 & 28 & 9.5 & 64.9 & 1.6\\
L19 & 82.8403.551 &   5 31 59.7   &   -69 19 51.5 &  19.8 & 0.15 &
2003-05-20 &  30 & 8.0 & 1.7 & -24.1 & 2004-07-11 & 29 & 11.6 & 26.1 & 1.1\\
L20 & 48.2620.2719 & 4 56 14.3 & -67 39 9.0 & 19.3 & 0.26 & 2002-12-01 &
 33 & 9.6 & 1.7 & 174.3 & 2005-06-04 & 31 & 12.3 & -1.1 & 2.5\\
L21 & 5.4892.1971  &  5 10 32.5   &   -69 27 15.5 &  18.9 & 1.58 &
2003-06-04 & 40 & 6.7 & 1.7 & -4.9 & 2004-07-18 & 35 & 9.9 & 38.0 & 1.1\\
\enddata
\tablecomments{The $V$ magnitudes
quoted here are from our $HST$ data and differ from the values in 
Geha \etal (2003) because 1) the MACHO resolution is much worse than 
that of the HRC and some of the sources were considerably blended and 
2) the QSOs have some intrinsic variability.  Epoch 1 
has program ID 9046 (Cycle 11) and epoch 2 has program ID 10130 (Cycle 13). 
ORIENTAT is the position angle on the sky of the detector $y$-axis (in 
degrees east of north). Redshifts are from Geha \etal (2003).}
\end{deluxetable}


\begin{deluxetable}{lcccccccccc}
\tabletypesize{\tiny}
\rotate
\tablewidth{0pt}
\tablecolumns{11}
\tablecaption{Results}
\tablehead{
\colhead{ID}  & \colhead{$N_{\rm{sources}}$} & 
\colhead{$N_{\rm{master}}$} & \colhead{$N_{\rm{used}}$\tablenotemark{a}} &
\multicolumn{4}{c}{PM of field as observed} & 
\multicolumn{2}{c}{LMC PM(CM) estimate} & \colhead{Used?}}
\startdata
 &  &     &  & $\mu_N$ & $\mu_W$ & $\delta\mu_N$ & $\delta\mu_W$ &
 $\mu_N$ & $\mu_W$ &  \\
   &   &    &    & ($\masyr$)  & ($\masyr$)  & ($\masyr$)  & ($\masyr$)  & 
($\masyr$)  & ($\masyr$)  &  \\
L1 & 85 & 35 & 15 & 0.800 & -1.162 & 0.089  & 0.121 & 0.550 & -1.334 & 0\\
L2 & 70 & 35 & 10 & -0.274  & - 2.220  &   0.067 &  0.070 & 0.238 &-2.098 & 0\\
L3 & 298 & 109 & 51 & 0.413 & -1.976 &    0.061  & 0.082 & 0.553 & -1.969 & 1\\
L4 & 97 & 48 & 17 & 0.521 & -2.121 &   0.118 &   0.150 &  0.692 & -2.218 & 1\\
L5 & 240 & 84 & 27 & 0.102 & -2.054 & 0.097  & 0.086 &0.234 & -2.007 & 1 \\
L6 & 41 & 16 & 7 & 0.983 & -1.658 &0.298 & 0.191 & 0.543 & -1.858 & 0\\ 
L7 & 622 & 269 & 116 & 0.426 & -2.068 &  0.064 &  0.073 & 0.543 & -2.045 & 1\\
L8 & 135 & 49 & 20 & -0.050 & -1.961 & 0.058  &  0.081 & 0.329 & -2.018 & 1\\
L9 & 213 & 81 & 42 & -0.047 & -2.023 &  0.084 &  0.077 & 0.406 & -1.967 & 1\\
L10 & 214 & 52 & 38 & 0.600 &  -1.934 &   0.087 &  0.099 & 0.337 & -2.014 & 1\\
L11 & 59 & 22 & 8 &  1.186 & -1.089 & 0.187 & 0.352 & 0.827 & -1.305 & 0\\
L12 & 153 & 58 & 20 & -0.024 & -2.447  & 0.183  &   0.169 & 0.338 & -2.478 & 1\\
L13 & 426 & 163 & 68 & 1.168  & -2.455  &  0.078 & 0.134 & 1.230 & -2.441 & 0\\
L14 & 97 & 34 & 18 & 0.136   &  -1.837 &  0.225  &   0.174 & 0.484 & -1.874 & 1\\
L15 & 340 & 147 & 46 & 0.573 & -2.738  &  0.217  &  0.173 & 0.719 & -2.704 & 1\\
L16 & 95 & 33 & 8 & 0.277 & -1.525  &  0.117  &  0.141 & 0.609 & -1.597 & 0\\
L17 & 59 & 25 & 5 & 1.004 & -2.313 & 0.185 & 0.289 &  1.067 & -2.449 & 0\\
L18 & 100 & 39 & 12 & 0.914 &  -1.779 &  0.128 &   0.133 & 0.897 & -1.908 & 0\\
L19 & 486 & 163 & 89 & 0.844 & -1.735 &  0.208 &   0.163 & 0.808 & -1.744 & 1\\
L20 & 178 & 44 & 25 & 0.005 & -1.813 & 0.116 & 0.071 & 0.441 & -1.833 & 1\\
L21 & 498 & 210 & 113 & 0.054 & -2.464 &  0.122 & 0.115 & 0.254 & -2.434 & 1\\
\enddata
\tablenotetext{a}{$N_{\rm{sources}}$ refers to the number of real sources 
(detected in at least half of the images). $N_{\rm{master}}$ refers to the 
number or sources in the master-list, i.e. detected in every image in 
every epoch. $N_{\rm{used}}$ refers to the number of sources that are used in the 
final linear transformations after the PM and $\delta$PM cuts. Columns 5-8 contain 
the PM estimates and their errors for each field.
 Columns 9 \& 10 contain the 
PM estimates for the LMC center of mass, which includes corrections for the viewing 
perspective and rotation effects discussed in \S4.1. The last column notes 
if the particular field was used in our 
final estimate of the center of mass motion of the LMC (Equation~(\ref{PMfinal})).} 
\end{deluxetable}

\end{document}